\documentclass[10pt]{article}

\usepackage{amsmath,amsmath,amssymb}
\usepackage{mathrsfs}
\usepackage{multirow}
\usepackage{tabularray}
\usepackage{adjustbox}
\usepackage{graphicx}
\usepackage{authblk}
\usepackage{indentfirst}
\usepackage{multicol}  
\usepackage{tabu}
\usepackage{tabularx}
\usepackage{url}
\usepackage{fancyhdr}
\usepackage[numbers]{natbib}
\usepackage{lineno}
\usepackage{makecell}
\usepackage{fancyhdr}
\usepackage{colortbl}
\usepackage{caption}
\usepackage{booktabs}
\usepackage{tikz-cd}
\usepackage{tikz}
\usepackage{float}
\usepackage[margin=1in]{geometry} 
\usepackage{adjustbox}
\usepackage{booktabs}

\usepackage{setspace} 

\usepackage{hyperref}
\hypersetup{colorlinks=true,linkcolor=blue,anchorcolor=blue,citecolor=blue}

\newtheorem{theorem}{Theorem}
\newtheorem{definition}{Definition}

\newtheorem{remark}{Remark}

\newtheorem{example}{Example}

\title{AlphaFunctor: Bridging The Gap Between Protein Function Annotation and Property Prediction
}

\author[1,2]{Xiang Liu} 
\author[3,4]{Anna E. Yee}
\author[3]{Josh V. Vermaas}
\author[5]{Daniel R. Woldring}
\author[1,2,6]{Guo-Wei Wei \thanks{Corresponding author: guowei.wei@uga.edu}}
\affil[1]{Department of Mathematics, Michigan State University, East Lansing, MI, 48824, USA}
\affil[2]{Department of Mathematics, University of Georgia, Athens, GA 30602, USA}
\affil[3]{Department of Biochemistry and Molecular Biology and MSU-DOE Plant Research Laboratory, Michigan State University, East Lansing, MI, 48824, USA} 
\affil[4]{Department of Mathematics, California State Polytechnic University, Pomona, CA, 91768, USA}
\affil[5]{Department of Chemical Engineering and Materials Science, Michigan State University, East Lansing, MI, 48824, USA} 
\affil[6]{Department of Biochemistry and Molecular Biology, University of Georgia, Athens, GA 30602, USA}
\date{}

\begin{document}
\maketitle

\paragraph{Abstract}

The fundamental relationship among protein sequence, structure, function, and physicochemical properties is a central principle in biology. While in principle protein function and properties should be able to be derived directly from protein sequence, in practice protein function and property prediction methods have been designed around specific datasets and specific property or function subsets,  
leading to an enormous gap between function annotation and property prediction. To address these challenges, we introduce AlphaFunctor, a category theory based foundation model-like platform to bridge the gap between protein function annotation and property prediction. 
Based on the hypothesis that protein function and properties can be directly derived from protein sequence, AlphaFunctor predicts protein functions as represented by Gene Ontology terms directly from sequence. Using these function predictions, AlphaFunctor further maps protein functions using topological spectral theory, path-complex neural networks, and protein domain analysis onto downstream property prediction. 
AlphaFunctor is (pre)trained in nearly 0.6 million protein function data points to deliver the state-of-the-art protein function annotation on three benchmark datasets. Without task-specific network redesign, AlphaFunctor maps qualitative protein
function annotation to various qualitative and quantitative protein property predictions, outperforming other dataset-specific and task-specific competing predictors.

  
\newpage
	 
\section{Introduction}
Proteins are fundamental biomolecules that carry out a wide range of functions based on their specific amino acid sequence. The biological functions within an organism for a given protein vary based on their catalytic activity, interactions with other biomolecules, and physicochemical properties. Despite how important accurate protein characterization is for elucidating biological mechanisms and developing therapeutic strategies for complex diseases \cite{eisenberg2000protein}, experimentally determining or computationally predicting protein function has remained technically challenging. The advent of high-throughput sequencing technology has led to an explosive number of protein sequences. However, fewer than 0.5\% of them are annotated by gene ontology (GO) terms \cite{gligorijevic2021structure}. GO terms exist for molecular functions (MF) such as enzymatic activity, cellular component (CC) to localize proteins within the cell, and biological process (BP) to connect proteins to phenotypes \cite{gene2019gene}. The largely uncharachterized proteome creates an urgent need for fast and accurate computational methods for protein function annotation and protein property prediction. Due to the high cost and time-consuming nature of traditional experimental methods, as well as the limited accuracy of homology-based approaches, machine learning methods have become a major driving force in protein function annotation and property prediction \cite{radivojac2013large}.

Machine learning approaches for the prediction of protein function and property can be broadly categorized into four groups: sequence-based methods, structure-based methods, protein-protein interaction (PPI)-based methods, and hybrid approaches. 
Sequence-based methods typically use amino acid composition, sequence motifs, or embeddings derived from pretrained protein language models to characterize proteins, such as DeepGOPlus~\cite{kulmanov2020deepgoplus}, ATGO~\cite{zhu2022integrating}, and TALE~\cite{cao2021tale}. Structure-based methods leverage three-dimensional protein structures and commonly employ graph neural networks for prediction, such as DeepFRI~\cite{gligorijevic2021structure}, GAT-GO~\cite{lai2022accurate}, and DPFunc~\cite{wang2025dpfunc}. PPI-based methods reply on the principle that interacting proteins often share similar functions, such as DeepGraphGO~\cite{you2021deepgraphgo}, Mashup~\cite{cho2016compact}, and GeneMANIA~\cite{mostafavi2008genemania}. In addition, hybrid models refer to methods using multiple sources of information, including structural, sequence, PPI, and biomedical literature features, such as Gostruct~\cite{kahanda2017gostruct} and GOBeacon~\cite{lin2025gobeacon}. 

The Critical Assessment of Functional Annotation (CAFA) challenges have shown that the best performing methods typically integrate these complementary perspectives within unified frameworks \cite{jiang2016expanded,zhou2019cafa}. However, many of these methods are available only for limited subsets of proteins or require substantial computational resources. Consequently, models that can learn directly from sequence information alone represent the most broadly applicable solution to the rapidly expanding protein sequence space. However, existing sequence-based methods have substantial limitations. First, most methods rely primarily on amino acid composition and local motif features, ignoring higher-order and non-local interactions among residues and subsequences. Secondly, current methods are designed for specific datasets,  specific biological tasks, and specific subsets of protein functions and/or protein properties. This limts model transferability to other biological datasets, labels, and tasks. Nevertheless, diverse protein tasks are ultimately governed by the shared biochemical and biophysical principles encoded within protein sequences. Therefore, there is a pressing need to develop innovative models to bridge the gap between protein function annotation and property predictions.  
 
Topological deep learning (TDL) \cite{cang2017topologynet}  provides a powerful strategy to address these challenges. One type of TDL integrates deep neural networks and topological features extracted by using topological data analysis (TDA) tools, such as persistent homology \cite{carlsson2009topology, edelsbrunner2002topological} and persistent Laplacians \cite{wang2020persistent}, to capture higher-order interactions and global structural patterns. For example,  topological sequence analysis (TSA)   maps (discrete) biomolecular sequences to topological or algebraic invariants \cite{hozumi2024revealing,suwayyid2025cakl} to facilitate TDL. 
The other type of TDL directly constructs topological neural networks, such as simplicial neural networks \cite{ebli2020simplicial} and path-complex neural networks \cite{li2024path}. 
Despite these advances, it remains unclear how to apply such topological frameworks to protein sequence modeling.

In this work, we introduce AlphaFunctor, a category theory-based framework to bridge the gap between protein function annotation and property prediction. AlphaFunctor integrates three complementary modalities, namely topological spectral theory, algebraic topology, and protein domain analysis, to map protein sequences to functions and to  properties. The topological spectral theory employs persistent Laplacians to construct a filtered simplicial complex from subsequence motifs and derives Laplacian embeddings that encode the topological structure and geometric connectivity of the high-order interactions among different subsequences. The algebraic topology models protein sequences as collections of $k$-paths through the framework of path complexes to capture higher-order relations among amino acid residues. Domain analysis extracts critical functional domain information. These complementary representations are jointly learned through a cross-attention mechanism that facilitates information exchange between branches, followed by feature fusion for prediction.

AlphaFunctor is built on the hypothesis that protein properties, such as binding affinity, stability, solubility, mutational response, cellular location, protein-protein interaction (PPI), etc., can be derived from protein function annotations. Conceptually, protein functions are modeled as objects in a biological ontology category, while protein properties are modeled as objects in a physical-property category, and learning or inference is designed as a functor between them. AlphaFunctor is pre-trained with qualitative protein function annotation, but predicts both qualitative  and quantitative protein properties.

Specifically, we first train and evaluate AlphaFunctor on three protein function benchmark datasets, CAFA3\cite{zhou2019cafa}, CAFA4, and the PDB dataset from DeepFRI\cite{gligorijevic2021structure}, across cellular component (CC), biological process (BP), and molecular function (MF) prediction tasks. AlphaFunctor consistently outperforms existing methods across all datasets and tasks, establishing itself as the top-performing framework for protein function prediction from sequences. 
Without task-specific network redesign, AlphaFunctor maps protein functions to diverse protein properties, including subcellular localization classification, predictions of mutation-induced protein solubility changes and protein--protein binding affinity changes,  protein--protein interaction classification and binding affinity prediction, and cyclic peptide function prediction. Evaluated on commonly used benchmark datasets, AlphaFunctor consistently achieves highly competitive performance, indicating that it captures fundamental protein principles rather than merely task-specific patterns. These results establish AlphaFunctor as a generalizable foundation model-like framework for protein analysis and biological applications.

\section{Results}
\subsection{Overview of AlphaFunctor}
\begin{figure}[htp]
    \centering
    \includegraphics[width=1\linewidth]{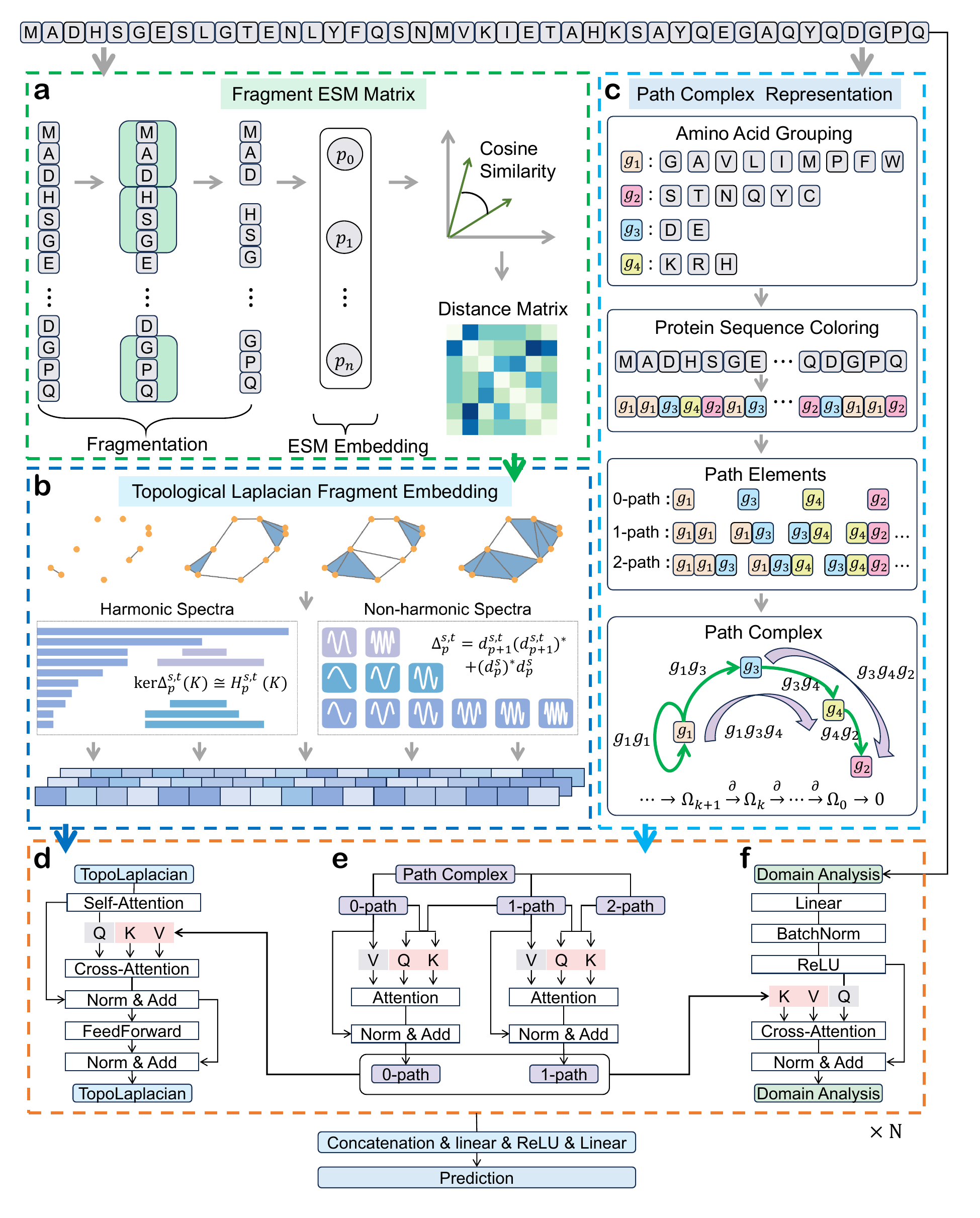}
    \vspace{-30pt}
    \caption{Model architecture of AlphaFunctor. AlphaFunctor employs three modalities, namely topological Laplacian, path complex, and domain analysis, to characterize protein sequences. In the topological Laplacian modality, the first step is to build a fragment evolutionary scale modeling (ESM) embedding and distance matrix (a). The persistent Laplacian (refer to Definition 7 and Theorem 1 of the Supplementary Information) is then computed from this distance matrix and organized into sequence-like embeddings (b), which are subsequently processed by a standard Transformer module (d). In the path complex modality, a path complex (refer to Equation (5) of the Supplementary Information ) is constructed from the protein sequence (c) and processed using a path complex neural network for feature extraction (e). In the domain modality, protein domain information is obtained via InterProScan\cite{jones2014interproscan} and represented as a binary vector. The vector is then fed into a multilayer perceptron (f). To effectively integrate these heterogeneous feature modalities, a cross-attention mechanism is applied within each layer. After several layers of updates, the outputs from all three modalities are concatenated and passed through a two-layer MLP for the prediction task.}
    \label{fig:model}
\end{figure}
AlphaFunctor integrates deep learning with features from three complementary perspectives, including topological Laplacian features, path complex features, and protein domain information, to characterize protein sequences. Specifically, three dedicated modalities corresponding to these feature types are designed to provide a comprehensive and robust protein representation. In the topological Laplacian modality, a fragment evolutionary scale modeling (ESM) embedding and distance matrix are built first (Figure \ref{fig:model}a). The distance matrix and embedding are used to compute the persistent Laplacian which is reorganized into a sequence-format embedding, analogous to inputs in natural language processing (Figure \ref{fig:model}b). This sequence embedding is subsequently processed using standard Transformer layers (Figure \ref{fig:model}d). In the path complex modality, a path complex is constructed from the protein sequence, where each path encodes a specific subsequence pattern within the protein (Figure \ref{fig:model}c). The resulting complex is then fed into a path complex neural network for feature extraction (Figure \ref{fig:model}e). In the domain modality, protein domain information is obtained using Interproscan \cite{jones2014interproscan} and represented as a binary vector, which is then processed by a simple multilayer perceptron (MLP) block (Figure \ref{fig:model}f). To effectively integrate these heterogeneous features, the cross-attention mechanism is employed to facilitate interactions across different branches. Finally, the features from all three modalities are concatenated and passed through a two-layer MLP to perform the prediction task.

For the topological Laplacian modality, a fragment strategy is applied to the protein sequence to extract subsequences, where multiple fragmentation sizes are employed to capture both local and global sequence patterns. For each derived subsequence, the average ESMC \cite{esm2024cambrian} embedding of its constituent amino acids is used as its coordinate representation. The cosine distance is then adopted to characterize the semantic similarity between subsequences within the protein context. Based on this distance, a filtered simplicial complex, specifically the Vietoris-Rips complex \cite{vietoris1927hoheren}, is generated to provide a multiscale topological interaction representation among these subsequence patterns. Subsequently, the persistent Laplacian is computed from the complex to characterize both the topological and geometric properties of these subsequences in the protein context. The harmonic spectra of the persistent Laplacian characterize global topological structures, such as connected components, loops, and voids, while the non-harmonic spectra capture additional geometric connectivity properties. These harmonic and non-harmonic spectra are concatenated and aligned along the filtration axis to form a sequence-format embedding. As the persistent Laplacian is constructed from pairwise cosine distances between ESMC embeddings, we additionally include the mean residue embedding as a class token to incorporate global, absolute information. The resulting token sequence is then fed into standard Transformer layers. 

For the path complex modality, the 20 amino acid types are first categorized into groups according to specific biological rules. For example, they can be partitioned into five physicochemical groups, including hydrophobic, uncharged, positively-charged, negatively-charged, and special cases, or treated as 20 distinct groups based on amino acid type. The given protein is then encoded as a colored sequence by assigning each amino acid to its corresponding group. Next, contiguous subsequences are extracted to form paths of different dimensions. Specifically, every single amino acid corresponds to a 0-path, an adjacent pair of amino acids form a 1-path, and adjacent triplets form 2-paths. More generally, a $k$-path corresponds to a contiguous subsequence of length $k$+1, following earlier definitions for paths \cite{grigor2020path}. By collecting all these paths, a weighted path complex is derived, where each path is associated with a weight reflecting its frequency of occurrence in the whole sequence. This weighted path complex is then processed by a path complex neural network in which each $k$-path is updated by aggregating information from its $k$-path neighbors and $(k+1)$-path cofaces.

The deep learning architecture for each layer of AlphaFunctor consists of three modules corresponding to the three feature types. A Transformer module is used to process the persistent Laplacian embedding, a path complex neural network module is applied to the path complex embedding, and an MLP module is used to process the domain embedding. To effectively integrate these heterogeneous representations, the cross-attention mechanism is employed within each layer to enable interactions across different feature modalities. After multiple such layers, the outputs of the three modules are concatenated and passed through a two-layer MLP to produce the final prediction.

\subsection{Performance on Protein Function Annotation}
\begin{figure}[ht]
    \centering
    \includegraphics[width=1\linewidth]{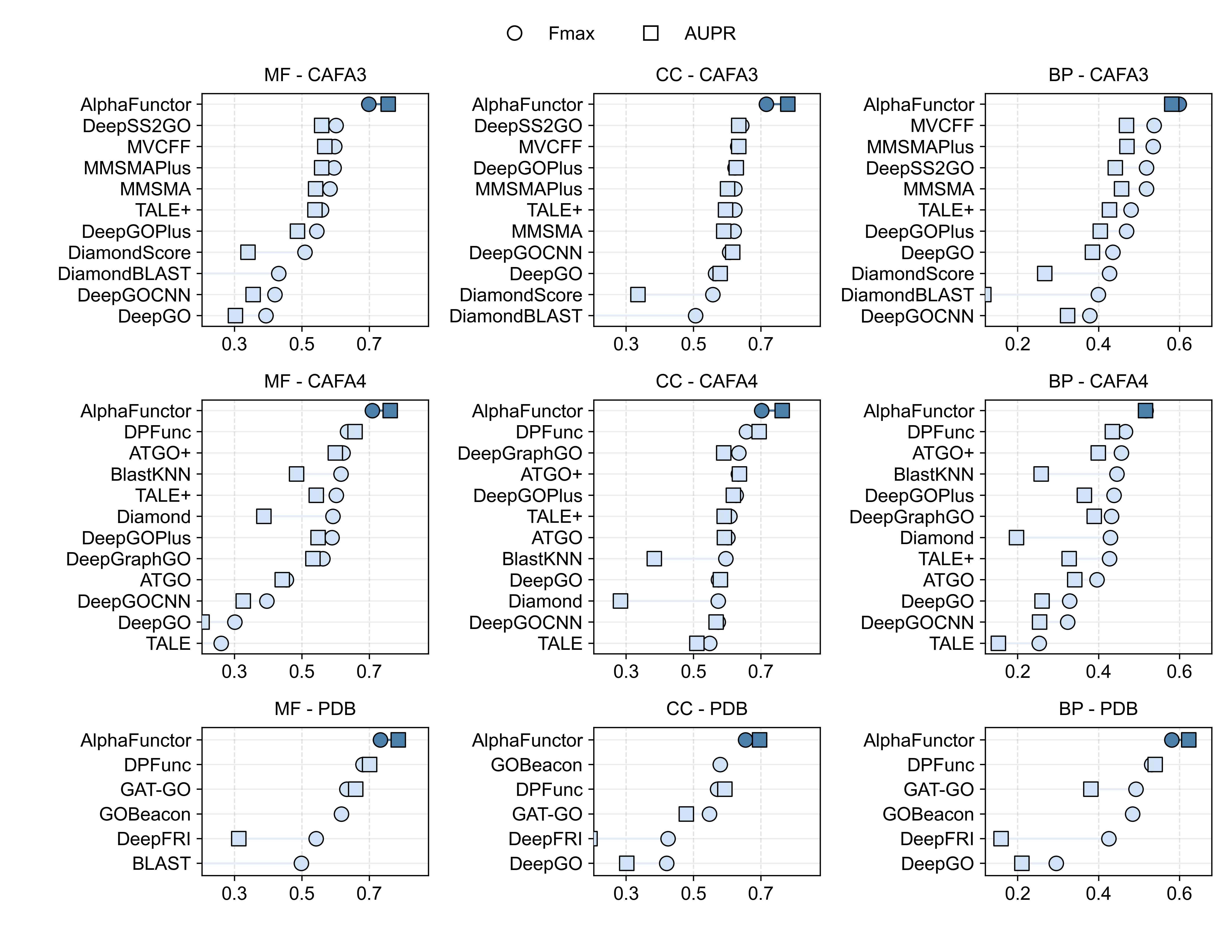}
    \caption{Performance comparison between AlphaFunctor and competing methods on CAFA3\cite{zhou2019cafa}, CAFA4, and PDB\cite{gligorijevic2021structure} datasets for protein function annotation. The first, second, and third columns correspond to molecular function (MF), cellular component (CC), and biological process (BP) tasks, respectively, while the first, second, and third rows correspond to CAFA3, CAFA4, and PDB datasets. AlphaFunctor achieves the strongest overall performance among the evaluated methods across the three tasks (CC, BP, and MF) and all three datasets, in terms of both Fmax (circle) and AUPR (square).}
    \label{fig:result-comparison}
\end{figure}
To evaluate the effectiveness of AlphaFunctor, we apply it to three benchmark datasets, CAFA3\cite{zhou2019cafa}, CAFA4, and PDB\cite{gligorijevic2021structure}, and compare it with existing methods, including BlastNKK~\cite{radivojac2013large}, Diamond~\cite{buchfink2015fast}, DeepGO~\cite{kulmanov2018deepgo}, DeepGOCNN~\cite{kulmanov2020deepgoplus}, DeepGOPlus~\cite{kulmanov2020deepgoplus}, TALE~\cite{cao2021tale}, TALE+~\cite{cao2021tale}, DeepGraphGO~\cite{you2021deepgraphgo}, DeepFRI~\cite{gligorijevic2021structure}, GAT-GO~\cite{lai2022accurate}, ATGO~\cite{zhu2022integrating}, MMSMA~\cite{wang2023mmsmaplus}, MMSMAPlus~\cite{wang2023mmsmaplus}, DeepSS2GO~\cite{song2024deepss2go}, DPFunc~\cite{wang2025dpfunc}, GOBeacon~\cite{lin2025gobeacon}, and MVCFF~\cite{yang2026multi}.
For fair comparison, we follow the existing training and testing splits of each dataset. Each experiment is repeated five times with different random seeds, and the averaged results are used as the final prediction of our model. Furthermore, to leverage the large number of samples in the SwissProt dataset, we first train AlphaFunctor on the SwissProt dataset \cite{boutet2007uniprotkb} and then use the learned weights to initialize the model for retraining on the three datasets via transfer learning. The results are shown in Figure~\ref{fig:result-comparison}, with detailed metric values provided in Supplementary Tables S3, S4, and S5.

As shown in Figure~\ref{fig:result-comparison}, AlphaFunctor consistently outperforms existing methods across all three datasets and functional tasks in terms of both Fmax and area under the precision-recall curve (AUPR).
For MF, AlphaFunctor achieves Fmax/AUPR scores of 0.698/0.756, 0.709/0.762, and 0.733/0.786 on CAFA3, CAFA4, and PDB datasets, respectively, yielding average improvements of 11.8\% in Fmax and 20.3\% in AUPR over the second-best methods.
For CC, AlphaFunctor obtains 0.716/0.780, 0.702/0.763, and 0.654/0.696 on the CAFA3, CAFA4, and PDB datasets, respectively, corresponding to average improvements of 10.4\% in Fmax and 16.7\% in AUPR.
For BP, AlphaFunctor achieves 0.599/0.581, 0.517/0.516, and 0.581/0.623, improving over the second-best methods by 10.6\% in Fmax and 19.4\% in AUPR on average.

Notably, although the second-best method varies across datasets and tasks, AlphaFunctor consistently achieves the best performance on all three datasets across all three GO categories.
These results demonstrate that AlphaFunctor learns effective protein representations and achieves strong overall performance in protein function prediction across the evaluated benchmarks.

\begin{figure}[htp]
    \centering
    \includegraphics[width=1\linewidth]{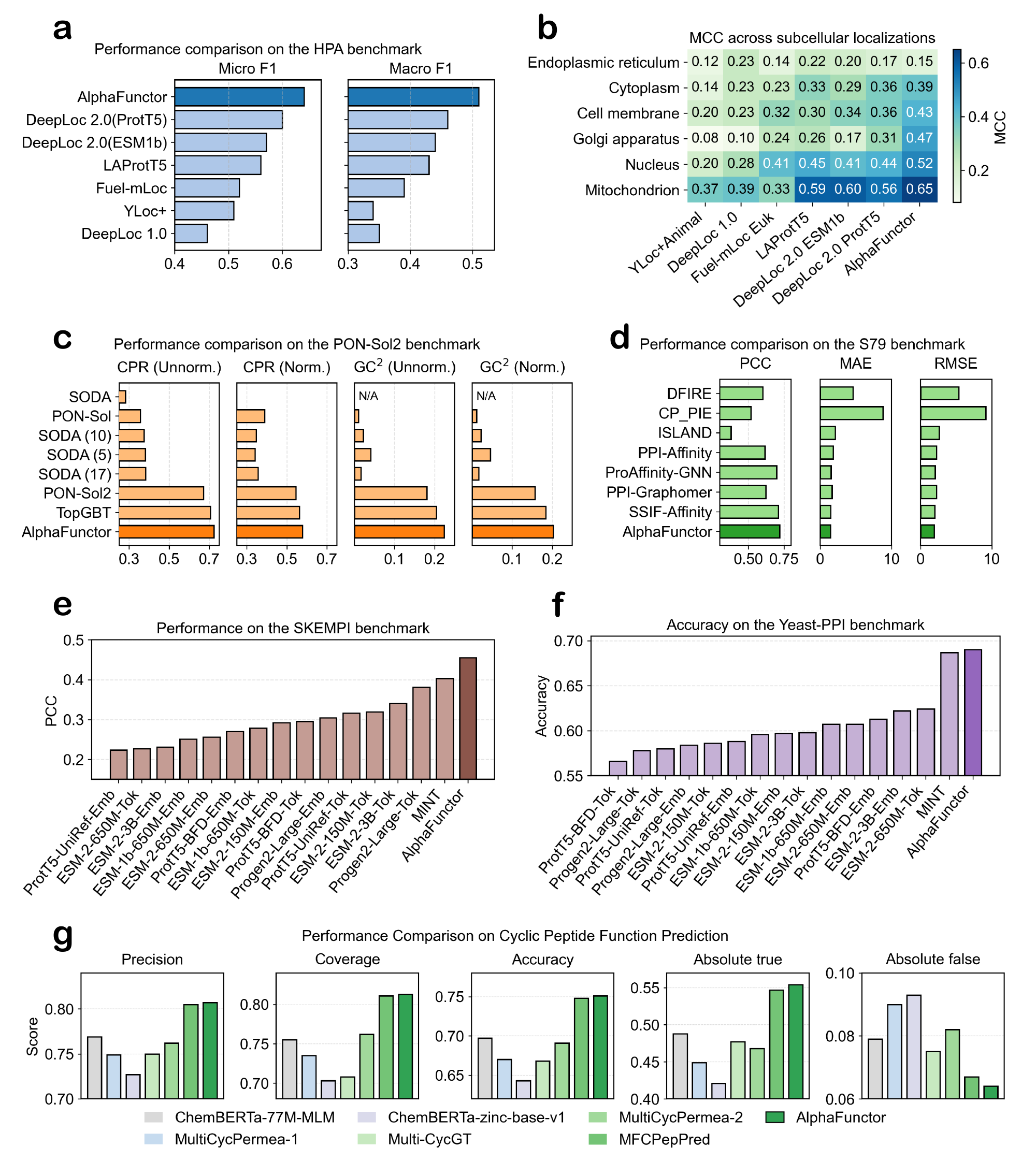}
    \caption{Comparison of AlphaFunctor and competing methods on protein property predictions. (a): Performance on the HPA dataset for protein subcellular localization prediction. (b): MCC on the HPA dataset across different subcellular localizations. (c): Performance on the PON-Sol2 dataset for mutation-induced protein solubility prediction. (d): Performance on the S79 dataset for protein--protein binding affinity prediction. (e): Performance on the SKEMPI dataset for predicting mutation-induced protein--protein binding affinity changes. (f): Performance on the Yeast-PPI dataset for protein--protein interaction binary classification. (g): Performance on the CyclicPeptide dataset for cyclic peptide function prediction.}
    \label{fig:generalization-result}
\end{figure}

\subsection{Function Annotation-enabled Subcellular Localization Prediction}
Protein subcellular localization is closely linked to function annotation and disease-associated mislocalization \cite{park2011protein}. We transferred AlphaFunctor, supervised-pretrained on the SwissProt function annotation dataset, to this task following the DeepLoc 2.0 benchmark \cite{thumuluri2022deeploc}. Specifically, the SwissProt Localization (SPL) dataset and the Human Protein Atlas (HPA) dataset were used as the training and test sets, respectively. 
We compared AlphaFunctor with existing benchmark methods, including YLoc+~\cite{briesemeister2010going}, DeepLoc 1.0~\cite{almagro2017deeploc}, Fuel-mLoc~\cite{wan2017fuel}, LAProT5~\cite{stark2021light}, and DeepLoc 2.0~\cite{thumuluri2022deeploc}. The results are shown in Figure \ref{fig:generalization-result}a and Figure \ref{fig:generalization-result}b, with detailed metric values provided in Supplementary Table S7. 
It can be seen that AlphaFunctor consistently outperforms existing methods in both micro F1 and macro F1 scores. Specifically, AlphaFunctor achieves micro-F1/macro-F1 of 0.64/0.51, compared with the second-best results of 0.60/0.46. Considering the Matthews correlation coefficient (MCC) for each localization category, AlphaFunctor also achieves the best performance in nearly all groups, except for the Endoplasmic reticulum category, which may be due to its relatively small sample size.
To further evaluate the robustness of our model, we assessed AlphaFunctor on the SPL dataset using five-fold cross-validation, strictly following the data splits adopted by DeepLoc 2.0 \cite{thumuluri2022deeploc}. AlphaFunctor consistently outperforms existing methods across all evaluation metrics. Specifically, AlphaFunctor achieves average microF1 and macroF1 scores of 0.81 and 0.77, respectively, significantly surpassing the second-best model, which attains 0.73 and 0.66. Detailed results are provided in Table S6 of the Supplementary Information.

\subsection{Function Annotation-enabled  Prediction of Protein Solubility Changes Upon Mutation}
Accurately predicting mutation-induced protein solubility changes is important as a single amino acid alteration can markedly alter solubility and consequently lead to disease~\cite{meulemans2010defining}. 
We transferred AlphaFunctor, supervised-pretrained on the SwissProt protein function dataset, to this task following the PON-Sol2 benchmark. AlphaFunctor was compared with existing benchmark methods, including PON-Sol~\cite{yang2016pon}, SODA~\cite{paladin2017soda}, PON-Sol2~\cite{yang2021pon}, and TopGBT~\cite{wee2024integration}. The results of our model and existing methods are presented in Figure \ref{fig:generalization-result}c. Detailed metric values are provided in Supplementary Table S8.
It can be seen that AlphaFunctor consistently outperformed existing methods on both correct prediction ratio (CPR) and generalized correlation (GC$^2$) and normalized metrics. Notably, AlphaFunctor achieved a CPR of 0.725, outperforming the second-best model's 0.707, and a GC$^2$ of 0.224 compared to the second-best result of 0.205. Furthermore, we categorize the test set into several groups based on mutation type, including charged, polar, hydrophobic, and special-case mutations, as well as alanine and non-alanine mutation groups. Our model achieves CPR scores of 0.79, 0.72, 0.68, 0.72, 0.79, and 0.72 for the charged, polar, hydrophobic, special-case, alanine, and non-alanine groups, respectively. The comparable performance across these groups suggests that AlphaFunctor is not strongly biased toward a specific mutation category. A more detailed analysis of model performance across each pair of mutation groups is provided in Supplementary Figure S3.

\subsection{Function Annotation-enabled Protein--Protein Interaction Classification}
We transferred AlphaFunctor, supervised-pretrained on the SwissProt protein function dataset, to protein--protein interaction classification.
We evaluated our model on the Yeast-PPI dataset used by MINT \cite{ullanat2026learning}, following the predefined data splits with a 40\% sequence identity threshold \cite{guo2008using}. Our model achieved the best performance, with an accuracy of 0.690, compared with 0.687 for the second-best model, MINT. In comparison, ESM2-3B (trained on approximately 45 million sequences) \cite{lin2023evolutionary}, ProtT5-BFD (trained on approximately 2.2 billion sequences) \cite{elnaggar2021prottrans}, and ProGen2-Large (trained on approximately 1 billion sequences) \cite{nijkamp2023progen2}, achieved accuracies of 0.622, 0.613, and 0.584, respectively. Detailed results are provided in Fig.~\ref{fig:generalization-result}f and Supplementary Table S10.

\subsection{Function Annotation-enabled Protein--Protein Binding Affinity Prediction}
We next transferred AlphaFunctor, supervised-pretrained on the SwissProt protein function dataset, to protein--protein binding affinity prediction. 
We evaluated our model on the S79~\cite{vangone2015contacts} dataset and compared it with existing benchmark methods, including DFIRE~\cite{rives2021biological}, CP\_PIE~\cite{ravikant2010pie}, ISLAND~\cite{abbasi2020island}, PPI-Affinity~\cite{romero2022ppi}, ProAffinity-GNN~\cite{zhou2024proaffinity}, PPI-Graphomer~\cite{xie2025ppi}, and SSIF-Affinity~\cite{xu2025ssif}. 
The results are shown in Figure \ref{fig:generalization-result}d, with detailed metric values provided in Supplementary Table S9.
Our model outperforms all existing methods in terms of PCC, RMSE, and MAE. Specifically, it achieves a PCC of 0.719, an RMSE of 1.93 kcal/mol, and an MAE of 1.45 kcal/mol, compared with the second-best model's score of 0.710, 2.00 kcal/mol, and 1.49 kcal/mol, respectively. 
Note that the existing top-performing method on S79, SSIF-Affinity, utilizes both sequence information and three-dimensional structural information of the protein--protein complex, whereas our model relies solely on sequence information.

\subsection{Function Annotation-enabled  Prediction of Protein--Protein Binding Affinity Changes Upon Mutation}
Protein--protein interactions and their mutation-induced perturbations are central to mutation-related diseases, drug design, and therapeutic intervention. We transferred AlphaFunctor, supervised-pretrained on the Swiss-Prot protein function dataset, to the prediction of mutation-induced protein--protein binding affinity changes.
We evaluated AlphaFunctor on the SKEMPI2 dataset \cite{jankauskaite2019skempi}, a manually curated database containing both single-point and multiple-point mutations. To assess the foundation-like properties of our model, we compared AlphaFunctor with the recently published MINT model and representative protein language models, including ESM, ProtT5, and ProGen2, using the same training and testing splits used by MINT. AlphaFunctor achieves a PCC of 0.455, substantially outperforming the second-best model, MINT, which achieves a PCC of 0.403. In comparison, ESM2-3B \cite{lin2023evolutionary}, ProtT5-BFD \cite{elnaggar2021prottrans}, and ProGen2-Large \cite{nijkamp2023progen2}, achieve PCC scores of 0.340, 0.295, and 0.381, respectively. Notably, MINT is specifically designed for protein--protein interactions and is self-supervised pretrained on 96 million protein--protein complexes, whereas AlphaFunctor is supervised-pretrained on fewer than 0.6 million single-protein sequences. Detailed results are shown in Figure \ref{fig:generalization-result}e and Supplementary Table S11.

\subsection{Function Annotation-enabled Cyclic Peptide Function Prediction}
Cyclic peptides are promising drug candidates because they often exhibit strong binding affinity, low toxicity, and high specificity toward target proteins. We transferred AlphaFunctor, supervised-pretrained on the Swiss-Prot protein function dataset, to cyclic peptide function prediction. 
AlphaFunctor was evaluated on the CyclicPepedia dataset and compared with existing benchmark methods, including ChemBERTa-77M-MLM \cite{chithrananda2020chemberta}, SMILES\_tokenized\_PubC \cite{wang2025multicycpermea}, ChemBERTa-zinc-base-v1 \cite{chithrananda2020chemberta}, Multi\_CycGT \cite{cao2024multi_cycgt}, MultiCycPermea \cite{wang2025multicycpermea}, and MFCPepPred \cite{lu2026mfcpedpred}. The results are shown in Figure \ref{fig:generalization-result}g and Supplementary Table S12. Our model consistently outperformed all competing methods across all evaluation metrics. Notably, ChemBERTa-77M-MLM, MultiCycPermea, and ChemBERTa-zinc-base-v1 are large language models pretrained on tens of millions of samples. In addition, the second-best model, MFCPepPred, uses not only SMILES information but also 3D structural information, whereas our model relies only on amino acid sequence information and still achieves the best performance.

\begin{figure}[ht]
    \centering
    \includegraphics[width=1\linewidth]{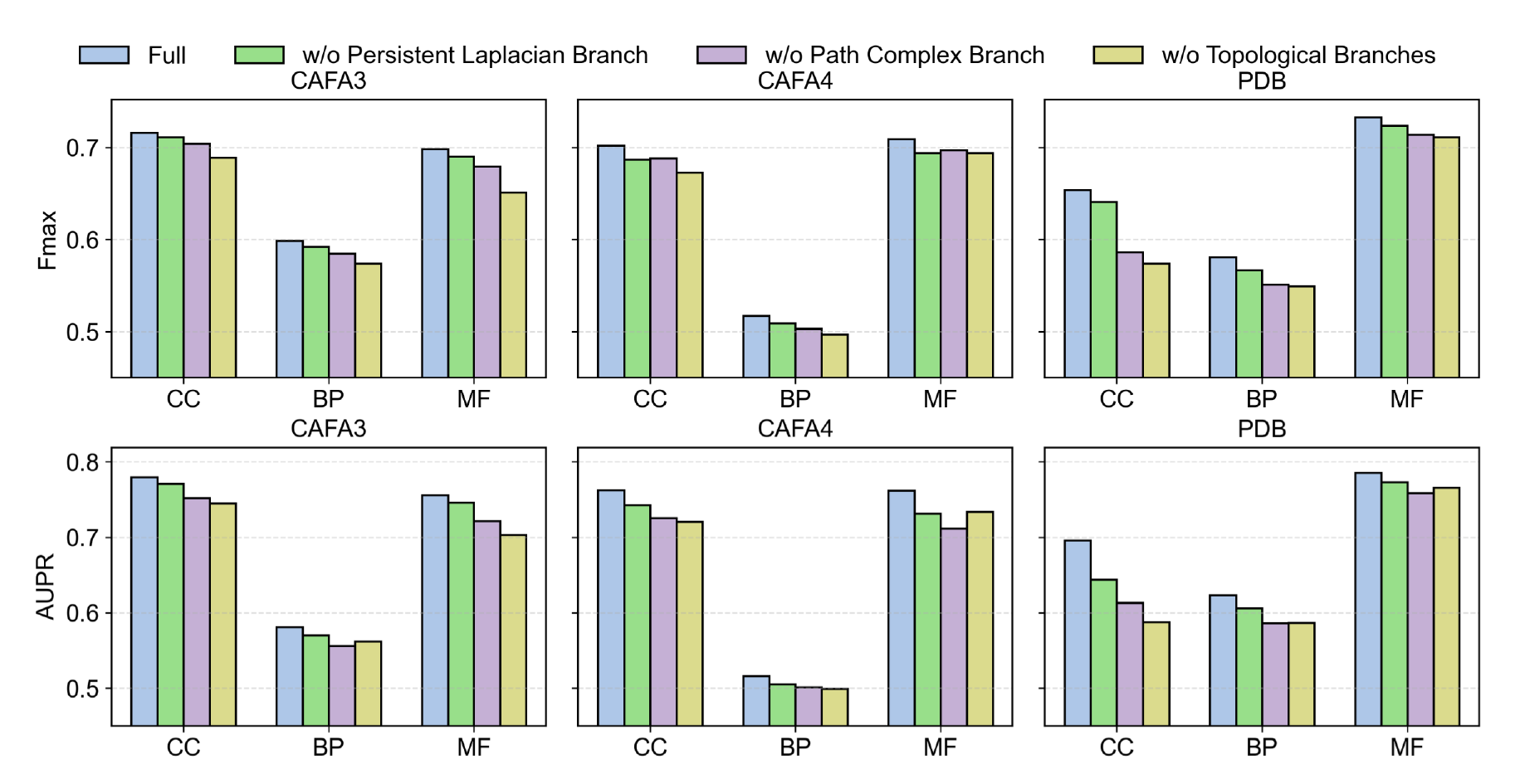}
    \caption{Ablation study on topological branches. The complete model is denoted as ``Full''. The variants excluding the persistent Laplacian branch, the path complex branch, and both topological branches are denoted as without (w/o) persistent Laplacian branch, w/o path complex branch, and w/o topological branches, respectively. The results indicate that the persistent Laplacian branch and the path complex branch provide complementary information, and each contributes positively to the overall performance.}
    \label{fig:ablation}
\end{figure}

\subsection{Ablation Study on Topological Branches}
AlphaFunctor incorporates two topological branches, namely the persistent Laplacian branch and the path complex branch. To evaluate their contributions, we conducted ablation studies on three benchmark datasets by removing each branch individually as well as both branches simultaneously. The results are presented in Figure \ref{fig:ablation}, tracking performance declines across all datasets.
Removing any topological component leads to a consistent degradation in performance, indicating that both branches provide meaningful contributions to predictive power. Moreover, the path complex branch generally yields a larger performance drop than the persistent Laplacian branch across most tasks and datasets, suggesting a relatively stronger contribution. Notably, removing both topological branches results in the most significant performance decline, demonstrating that the two branches capture complementary information.
Specifically, when both branches are removed, the Fmax/AUPR for the CC task decreases from 0.716/0.780 to 0.689/0.745 on CAFA3, from 0.702/0.763 to 0.673/0.721 on CAFA4, and from 0.654/0.696 to 0.574/0.588 on PDB. For the MF task, the Fmax/AUPR decreases from 0.698/0.756 to 0.651/0.703 on CAFA3, from 0.709/0.762 to 0.694/0.734 on CAFA4, and from 0.733/0.786 to 0.714/0.766 on PDB. For the BP task, the corresponding decreases are from 0.599/0.581 to 0.574/0.572 on CAFA3, from 0.517/0.516 to 0.497/0.499 on CAFA4, and from 0.581/0.623 to 0.551/0.587 on PDB.
These results suggest that the topological branches contribute most significantly to the CC task compared to MF and BP.

\section{Discussion}
This study demonstrates that AlphaFunctor can effectively predict multiple quantities of interest. AlphaFunctor not only achieves state-of-the-art performance on protein function annotation benchmarks, but also generalizes effectively to diverse property prediction tasks, including subcellular localization, mutation-induced solubility, protein--protein binding classification, protein--protein binding affinity, mutation-induced binding affinity change, and cyclic peptide function prediction. These results highlight the broad applicability of function-supervised topological representations for protein function and property prediction.

For protein function prediction, we further directly applied AlphaFunctor to the CAFA3\cite{zhou2019cafa}, CAFA4, and PDB\cite{gligorijevic2021structure} datasets without transfer learning. AlphaFunctor consistently and substantially outperformed existing methods across all three datasets and all three task types (Supplementary Table S3, Table S4, and Table S5).
To further assess the robustness of our framework, we performed five-fold cross-validation on SwissProt, with folds constructed using CD-HIT \cite{fu2012cd} to ensure at most 40\% sequence identity between folds. On average, AlphaFunctor achieved Fmax/AUPR scores of 0.868/0.907 for MF, 0.824/0.896 for CC, and 0.796/0.847 for BP.
We also stratified the sequences into different groups according to sequence length and species. AlphaFunctor achieved stable performance across both sequence-length groups and species groups (Section S4.1 and S4.2 of the Supplementary Information), so it appears to be widely applicable. In addition, we conducted an ablation study on sequence identity, showing that AlphaFunctor maintains robust performance under varying levels of sequence similarity (Section S4.3 of Supplementary Information). These results further demonstrate the superiority and robustness of AlphaFunctor for protein function prediction.

The one caveat to this is that prediction performance is strongest when comparing to SwissProt/Uniprot GO annotations.
GO annotations in Uniprot use only selected automatic annotation methods to populate annotations where there is no direct experimental evidence.
By contrast, QuickGO uses more automated annotation tools to make more logical but not experimentally validated connections between a protein and a GO annotation \cite{huntley2015goa}.
Since AlphaFunctor was trained on the smaller set of experimental annotations, it does not recover the full set of annotations in larger datasets like QuickGO \cite{Binns2009}.
Protein function predictions from AlphaFunctor can be compared for the full SwissProt set of proteins to GO annotations from either Uniprot or QuickGO.

When using Uniprot as the ground truth, AlphaFunctor catches almost all of the existing annotations, with novel predictions outnumbering missed predictions (Figure S3).
This suggests that AlphaFunctor is learning novel relationships that go beyond what is in the initial training set, making inferences about what protein features correspond to a specific GO annotation.
However, when using the more expansive QuickGO annotations as the ground truth, AlphaFunctor is missing many of the QuickGO determined through automatic annotations without experimental support or validation.
Dataset quality matters for AI training, and so we have defaulted to the most conservative ground truth during training to avoid spurious annotations that may lower accuracy in real world applications.

Moving beyond GO annotations alone, AlphaFunctor is a consistent high performer in predicting protein-protein interactions, as judged by the Yeast-PPI and SKEMPI datasets.
It also highlights that bigger training datasets may not always yield better results.
MINT is pretrained on 96 million sequences \cite{ullanat2026learning}, ESM2-3B on 45 million sequences\cite{lin2023evolutionary}, ProtT5-BFD on approximately 2.2 billion sequences\cite{elnaggar2021prottrans}, and ProGen2-Large on approximately 1 billion sequences\cite{nijkamp2023progen2}.
ProtT5-BFD \cite{elnaggar2021prottrans}, and ProGen2-Large \cite{nijkamp2023progen2}, compared with only 600,000 sequences for AlphaFunctor.
This highlights the importance for choosing an efficient representation for the problem, and suggests that our topological framework and protein representations learned through function-supervised training provide an effective and transferable foundation for diverse biological tasks.

We anticipate that AlphaFunctor will remain an evolving tool, with further improvement possibilities that remain to be explored. We anticipate that more advanced topological tools, such as a persistent sheaf Laplacian~\cite{wei2025persistent},  
may provide deeper insights into protein sequence organization.
We also intend to explore integrating 3D structural or dynamical information to further enhance the current sequence-based framework. Another promising direction is to make AlphaFunctor explicitly aware of the hierarchical nature of GO annotations: Gene Ontology (GO) terms naturally form hierarchical structures through is\_a and part\_of relationships~\cite{huntley2015goa}. Such prior knowledge could be incorporated into the loss function to encourage predictions that more strictly satisfy these hierarchical constraints, which warrants future exploration. All of these advances will be in the service of extending AlphaFunctor to broader biological applications.

\section{Methods}
\subsection{Datasets and Evaluation Protocols}
For the downstream task evaluations, we repeated each experiment five times with different random seeds and used the averaged predictions across the five runs as the final model output. Additional implementation details are provided in Section S3.3 of the Supplementary Information. Here, we summarize the datasets used in this study, with details provided in Supplementary Table S1.

For protein function prediction, we considered three benchmark datasets: CAFA3, CAFA4, and PDB. 
The CAFA3 dataset is derived from the CAFA3 challenge~\cite{zhou2019cafa} and has been widely used in previous studies~\cite{kulmanov2018deepgo,kulmanov2020deepgoplus,wang2023mmsmaplus,yang2026multi}. The training set consists of protein sequences with experimental annotations available before September 2016, whereas the test set contains sequences annotated between September 2016 and November 2017. In total, this dataset contains 677 MF GO terms, 551 CC GO terms, and 3,992 BP GO terms. 
The CAFA4 dataset was collected by the DPFunc model~\cite{wang2025dpfunc} from the CAFA4 Challenge. Specifically, the training set includes sequences released before May 2020, the validation set includes sequences released between May 2020 and April 2021, and the test set includes sequences released between May 2021 and April 2022. This dataset contains 6,166 MF GO terms, 2,548 CC GO terms, and 19,825 BP GO terms.
The PDB dataset was introduced by the DeepFRI model~\cite{gligorijevic2021structure} and was divided into training, validation, and test sets using CD-HIT~\cite{fu2012cd} at 40\% sequence identity. It contains 489 MF GO terms, 320 CC GO terms, and 1,943 BP GO terms.

For subcellular localization prediction, the SwissProt Localization (SPL) dataset and the Human Protein Atlas (HPA) dataset were used as the training and testing sets, respectively. These datasets were curated by the DeepLoc 2.0 model~\cite{thumuluri2022deeploc}. Specifically, the SPL dataset is derived from the UniProt release 2021\_03 and contains ten localization classes: Cytoplasm, Nucleus, Extracellular, Cell membrane, Mitochondrion, Plastid, Endoplasmic reticulum, Lysosome/Vacuole, Golgi apparatus, and Peroxisome.
The HPA dataset is derived from the Human Protein Atlas project and shares at most 30\% sequence identity with the entire SwissProt dataset, as determined using USEARCH v11.0.667 (32-bit)~\cite{edgar2010search}.

For mutation-induced protein solubility prediction, the PON-Sol2 dataset~\cite{yang2021pon} was employed. This dataset contains 6,328 mutation samples from 77 proteins and is divided into three categories: decreased solubility, increased solubility, and no change in solubility. We followed the original training and testing split of PON-Sol2, using 5,666 samples for training and 662 samples for testing.

For protein--protein binding affinity prediction, we followed the SSIF-Affinity model~\cite{xu2025ssif} and used the protein--protein binding affinity data from PDBBind-v2020 as the training set. Test samples appearing in the training set were removed to avoid data leakage. The training data were extracted from the PDBbind-v2020 database, as in previous studies~\cite{romero2022ppi,xu2025ssif}. PDBbind-v2020 contains 2,852 protein--protein complexes with experimentally measured binding affinities. Only complete entries with reported $K_d$, $K_i$, and $IC_{50}$ were selected. The benchmark test set S79 was used, which is derived from the Structure-based Protein--Protein Binding Affinity Benchmark database~\cite{kastritis2011structure}. For the binary protein--protein interaction classification task, we use the Yeast-PPI dataset curated by Guo et al.~\cite{guo2008using}, which is a balanced dataset of interacting and non-interacting samples. A 40\% sequence identity threshold is applied across the training, validation, and test sets to reduce redundancy.

For mutation-induced protein--protein binding affinity change prediction, we employed the SKEMPI2 dataset \cite{jankauskaite2019skempi}, a manually curated dataset containing both single-point and multi-point mutations. To mitigate the risk of information leakage from structurally similar complexes, we followed the MINT model and divided the dataset into three mutually exclusive folds based on protein--protein complexes. Each fold contains unique complexes that do not appear in the other folds.

For cyclic peptide function prediction, we employed the benchmark dataset used in MFCPepPred \cite{lu2026mfcpedpred}, which was curated from the CyclicPepedia database \cite{liu2024cyclicpepedia} and consists of cyclic peptides with experimentally validated biological functions. The dataset contains 1,322 cyclic peptides spanning 14 functional categories. Following the evaluation protocol of MFCPepPred, we performed five-fold cross-validation to assess model performance.

\subsection{Persistent Laplacian Representation of Molecular  Sequences}
\begin{figure}[ht]
    \centering
    \includegraphics[width=0.8\linewidth]{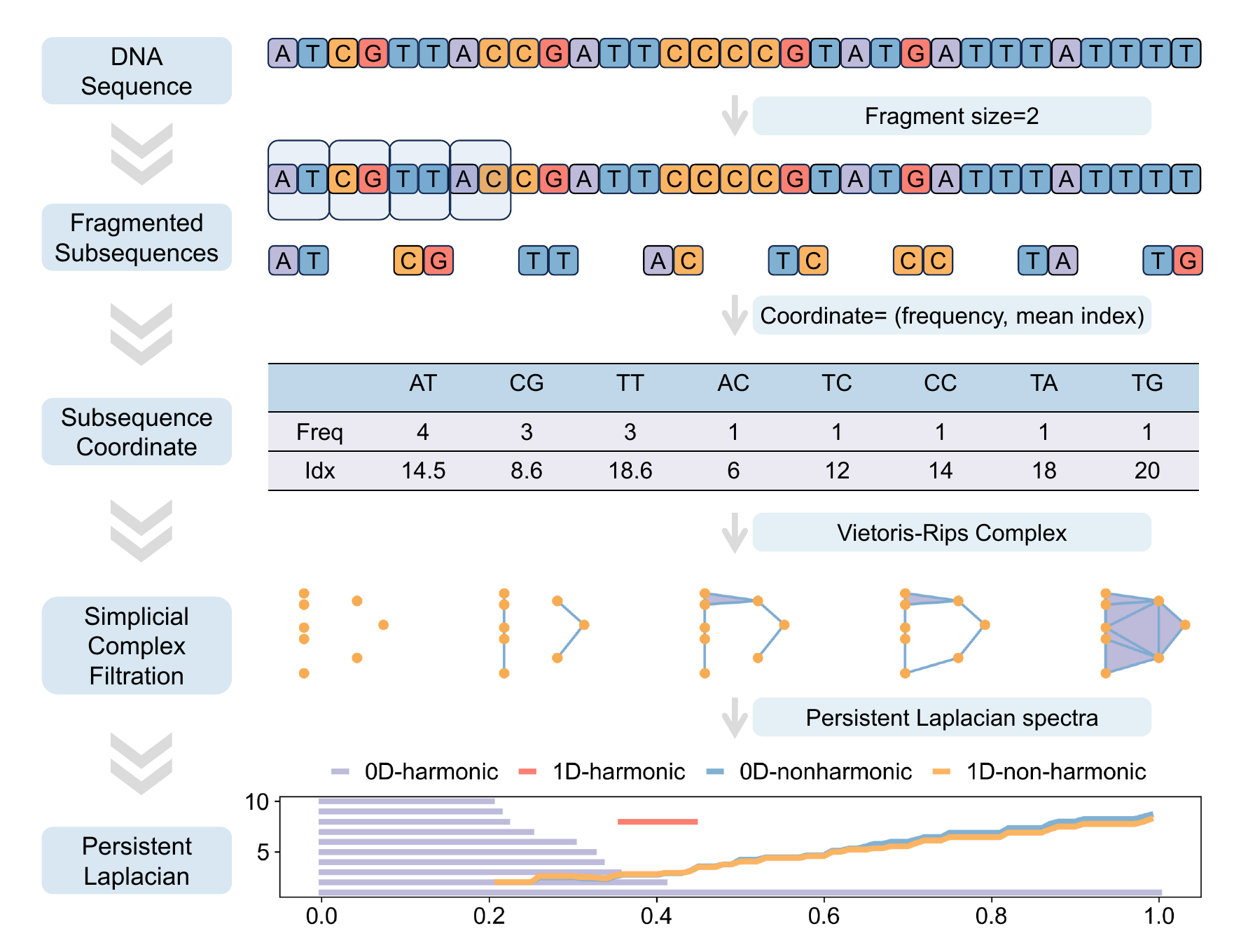}
    \caption{Illustration of the persistent Laplacian representation for a DNA sequence. For the given DNA sequence, the fragmentation method with a fragment size of 2 is first applied to extract a set of subsequences. Identical subsequence patterns are then merged, and each is assigned a coordinate pair (freq,idx), where freq denotes the frequency of occurrence of the subsequence within the entire sequence, and idx represents the average positional index of all its occurrences. Based on these coordinates, a Vietoris-Rips complex is constructed, yielding a sequence of nested simplicial complexes. The persistent Laplacian is then computed from this filtration. Subsequently, the harmonic and non-harmonic spectra of the persistent Laplacian are presented.  }
    \label{fig:pl}
\end{figure}

Persistent Laplacian theory \cite{wang2020persistent} is a spectral-theory generalization of persistent homology in topological data analysis. Theoretically, the persistent Laplacian not only captures topological information through its harmonic components, but also provides additional geometric and combinatorial information about the underlying space. A detailed description is provided in Section S2.1 of the Supplementary Information. Here, we briefly outline the procedure for constructing persistent Laplacians from sequence data.

Given a sequence $S=s_1s_2\cdots s_n$ of length $n$, we first employ a fragmentation strategy to partition the sequence into a collection of subsequences. Specifically, the original sequence is divided into equally sized fragments of a prescribed length. For example, when $S=abcd$ and the fragment size is 2, the resulting subsequences are $ab$ and $cd$.

Next, we assign a coordinate representation to each subsequence $S_i$. A simple choice is to use its positional index within the original sequence. Alternative representations include the frequency of occurrence of the subsequence within $S$. For protein sequences, more informative biological representations can also be employed, such as the average ESM embedding or other physicochemical descriptors, which provide biologically meaningful information for defining distances among subsequences.

After obtaining the coordinates of all subsequences, we compute the distance matrix $D$ among them. Standard choices include Euclidean distance, while other metrics such as cosine similarity distance \cite{nguyen2010cosine} can also be adopted. This distance matrix $D$ provides the foundation for applying mathematical tools from topology and geometry to analyze the original sequence data.

Based on the distance matrix $D$, a natural way to characterize interactions and higher-order structural patterns among subsequences is to construct a simplicial complex. Roughly speaking, a simplicial complex is a collection of simplices, including vertices, edges, triangles, and their higher-dimensional counterparts. These simplex components effectively capture both pairwise and many-body interactions among subsequences. Several constructions may be considered, such as the Vietoris-Rips complex \cite{vietoris1927hoheren}, Alpha complex \cite{edelsbrunner2011alpha}, or other graph-based complexes \cite{liu2022dowker}. Such constructions naturally generate a nested sequence of simplicial complexes
$$C_1\subset C_2\subset C_3\subset \cdots \subset C_n$$
where each complex $C_i$ corresponds to a specific scale parameter. This filtration provides a multiscale characterization of interactions among subsequences.

Finally, the persistent Laplacian is computed from this simplicial complex filtration. The harmonic spectra of the persistent Laplacian are equivalent to persistent homology information and characterize topological structures such as connected components, loops, and voids. In contrast, the non-harmonic spectra encode additional geometric connectivity information. These spectral features can be naturally organized as a sequential embedding along the filtration scale axis, making them well-suited for natural language processing models.

Since protein sequences involve 20 amino acid types, we illustrate our method using a DNA sequence involving nucleic acid types. As shown in Figure \ref{fig:pl}, the fragmentation method is applied first to the sequence to extract subsequences. Specifically, the fragment size is set to 2 so that all extracted subsequences have length 2. Next, relative coordinates are constructed for these subsequences. Identical subsequence patterns are merged and each resulting pattern is assigned a coordinate pair (freq, idx), where ``freq'' denotes the frequency of occurrence of the subsequence and ``idx'' represents the average positional index of all its occurrences. These coordinates are also normalized to the interval [0,1]. In this way, the original sequence data are transformed into a point cloud representation. 
On the basis of this point cloud, a Vietoris-Rips complex is constructed, yielding a sequence of nested simplicial complexes. Finally, the persistent Laplacian is computed from this simplicial complex filtration. The harmonic spectra of the persistent Laplacian characterize the topological structural information of the data, whereas the non-harmonic spectra encode additional geometric properties. 
As shown in Figure \ref{fig:pl}, there are ten 0D-harmonic bars, corresponding to the ten distinct subsequence patterns. The evolution of these bars along the $x$-axis reflects the changes in connected components among the subsequence patterns during the filtration process. The single 1D-harmonic bar corresponds to a loop structure formed during the simplicial complex filtration. Notably, the harmonic information remains unchanged after the filtration value reaches 0.5, whereas the non-harmonic spectral curves continue to increase gradually, which corresponds to the additional geometric information beyond that represented by the harmonic spectra.

\subsection{Path Complex Representation of Sequences}
Path complex theory was originally developed for directed graphs and is designed to provide a higher-order interaction analysis of the underlying graph structure. A rich mathematical framework has been established for path complexes   \cite{grigor2023homotopy}. This development offers powerful quantitative tools for practical applications. A brief description of path complex homology is provided in Section S2.2 of the Supplementary Information.

To construct the path complex representation of protein sequences, we first partition the 20 standard amino acids into groups. The most straightforward strategy is to treat the 20 amino acid types as 20 distinct groups. Alternatively, the grouping can be defined according to biologically meaningful criteria. For example, amino acids may be grouped as hydrophobic, polar uncharged, positively charged, and negatively charged based on physicochemical properties. Other grouping schemes can also be derived from side-chain structure or secondary-structure preference.
Having such a grouping rule, we transform the given protein sequence $S=s_1s_2\cdots s_n$ to a new sequence $T=t_1t_2\cdots t_n$, where $t_i$ denotes the group identity of amino acid $s_i$. This new sequence $T$ provides a simplified representation of the original sequence $S$ while preserving the specific biological characteristics encoded by the chosen grouping rule.

Next, we construct a path complex from this new sequence $T$. Specifically, each individual element $t_i$ is a 0-path, and each pair of adjacent elements $t_it_{i+1}$ is a 1-path. Generally, each consecutive subsequence $t_it_{i+1}\cdots t_{i+k}$ is a $k$-path. Collecting all such paths naturally yields a path complex. Generally speaking, a path complex $C$ is a collection of paths such that, for every path $p\in C$, the paths obtained by removing the first or last element of $p$ also belong to $C$. The construction adopted here automatically satisfies this condition. Note that some paths may occur multiple times within the sequence. Consequently, the resulting complex is more precisely a weighted path complex, in which each path is assigned a weight corresponding to its frequency of occurrence in the full sequence. This weighted path complex captures subsequence patterns of the original protein sequence under the chosen biological grouping rule, and provides a foundation for subsequent quantitative analysis using topological methods. Furthermore, it can be naturally employed as the input to a path complex neural network.

\subsection{AlphaFunctor Neural Network Layer}
In the neural network component of AlphaFunctor, each layer consists of three branches. The first branch is a standard Transformer layer for processing the persistent Laplacian embeddings. The second branch is a simple MLP for processing the domain features. The third branch is a path complex neural network for learning from the path complex representation. Cross-attention is employed within each layer to facilitate communication among these three types of features. After several such layers, the outputs of the three branches are concatenated and fed into a two-layer MLP for the final prediction task.

To introduce the path complex neural network, we first need several notations. Let $C$ be a path complex, and let $\sigma_n=s_ks_{k+1}\cdots s_{k+n}$ and $\tau_n=t_kt_{k+1}\cdots t_{k+n}$ be two $n$-paths. We say that $\sigma_n$ is a neighbor of $\tau_n$ if $s_i=t_{i-1}$ for $k+1\leq i\leq k+n$. If $\sigma_n$ is a neighbor of $\tau_n$, then the $(n+1)$-path $s_ks_{k+1}\cdots s_{k+n}t_{k+n}$, denoted by $(\sigma_n,\tau_n)$, is called the coface pointing from $\sigma_n$ to $\tau_n$.

In the path complex neural network, each path representation is updated by aggregating information from its neighbors and cofaces through a message-passing mechanism. Formally, let $\sigma_n\in C$ be an $n$-path. Its update is defined as
$$h_{\sigma_n}'=\sum_{\tau_n\in \mathcal{N}(\sigma_n)\cup\{\sigma_n\}}\alpha_{\tau_n,\sigma_n}h_{\tau_n}$$
where $h_{\sigma_n}'$ denotes the updated representation of $\sigma_n$ after one message-passing layer, $\mathcal{N}(\sigma_n)$ is the set of all neighbors of $\sigma_n$, $\alpha_{\tau_n,\sigma_n}$ is the attention coefficient from $\tau_n$ to $\sigma_n$, and $h_{\tau_n}$ is the representation of the neighboring path $\tau_n$.
The attention coefficient $\alpha_{\tau_n,\sigma_n}$ depends on the representations of $\sigma_n$, $\tau_n$, and their coface $(\tau_n,\sigma_n)$, and is computed as
$$\alpha_{\tau_n,\sigma_n}=\frac{\exp{[\text{LeakyReLU}(\theta_s(h_{\tau_n})+\theta_t(h_{\sigma_n})+\theta_c(h_{\tau_n,\sigma_n}))]}}{\sum_{\beta_n\in\mathcal{N}(\sigma_n)\cup\{\sigma_n\}}\exp{[\text{LeakyReLU}(\theta_s(h_{\beta_n})+\theta_t(h_{\sigma_n})+\theta_c(h_{(\beta_n,\sigma_n)}))]}}$$
where $\theta_s$, $\theta_t$, and $\theta_c$ are learnable linear maps corresponding to the source path, target path, and the connecting coface, respectively. Details of all hyperparameters, evaluation metrics across all tasks, and the adaptation of AlphaFunctor to diverse protein property prediction tasks are provided in Sections S3.1, S3.2, and S3.3 of the Supplementary Information, respectively.

\section*{Acknowledgment}
This work was supported in part by NIH grants R01AI164266 and  R35GM148196 to GW, Michigan State University Research Foundation, The University of Georgia, and Georgia Research Alliance.
This material is based in part upon work supported by the Defense Advanced Research Projects Agency (DARPA) under Agreement No. HR00112969E087.
A.E.Y was supported by the ACRES REU through NSF Grant 2349002.
Research infrastructure is supported in part by the U.S. Department of Energy, Office of Science, Office of Basic Energy Sciences, United States Department of Energy under Award Number DE-FG02-91ER20021.
The views and conclusions contained in this document are those of the authors and should not be interpreted as representing the official policies, either expressed or implied, of the U.S. Government.
This work was supported in part through computational resources and services provided by the Institute for Cyber-Enabled Research at Michigan State University.

\section*{Declaration of Interests}
The authors declare no competing interests.

\bibliographystyle{unsrtnat}
\bibliography{ref }

@article{Binns2009,
  title = {{{QuickGO}}: A Web-Based Tool for {{Gene Ontology}} Searching},
  shorttitle = {{{QuickGO}}},
  author = {Binns, David and Dimmer, Emily and Huntley, Rachael and Barrell, Daniel and O'Donovan, Claire and Apweiler, Rolf},
  year = 2009,
  month = nov,
  journal = {Bioinformatics},
  volume = {25},
  number = {22},
  pages = {3045--3046},
  issn = {1367-4811, 1367-4803},
  doi = {10.1093/bioinformatics/btp536},
  urldate = {2026-07-07},
  copyright = {http://creativecommons.org/licenses/by-nc/2.0/uk/},
  langid = {english}
}

@article{ebli2020simplicial,
  title={Simplicial neural networks},
  author={Ebli, Stefania and Defferrard, Micha{\"e}l and Spreemann, Gard},
  journal={arXiv preprint arXiv:2010.03633},
  year={2020}
}

@article{carlsson2009topology,
  title={Topology and data},
  author={Carlsson, Gunnar},
  journal={Bulletin of the American Mathematical Society},
  volume={46},
  number={2},
  pages={255--308},
  year={2009}
}

@article{cang2017topologynet,
  title={TopologyNet: Topology based deep convolutional and multi-task neural networks for biomolecular property predictions},
  author={Cang, Zixuan and Wei, Guo-Wei},
  journal={PLoS computational biology},
  volume={13},
  number={7},
  pages={e1005690},
  year={2017},
  publisher={Public Library of Science San Francisco, CA USA}
}

@article{wang2020persistent,
  title={Persistent spectral graph},
  author={Wang, Rui and Nguyen, Duc Duy and Wei, Guo-Wei},
  journal={International journal for numerical methods in biomedical engineering},
  volume={36},
  number={9},
  pages={e3376},
  year={2020},
  publisher={Wiley Online Library}
}

@article{lin2023evolutionary,
  title={Evolutionary-scale prediction of atomic-level protein structure with a language model},
  author={Lin, Zeming and Akin, Halil and Rao, Roshan and Hie, Brian and Zhu, Zhongkai and Lu, Wenting and Smetanin, Nikita and Verkuil, Robert and Kabeli, Ori and Shmueli, Yaniv and others},
  journal={Science},
  volume={379},
  number={6637},
  pages={1123--1130},
  year={2023},
  publisher={American Association for the Advancement of Science}
}

@article{wei2025persistent,
  title={Persistent Topological {Laplacians--a Survey}},
  author={Wei, Xiaoqi and Wei, Guo-Wei},
  journal={Mathematics},
  volume={13},
  number={2},
  pages={208},
year={2025}
}

@article{vietoris1927hoheren,
  title={{\"U}ber den h{\"o}heren Zusammenhang kompakter R{\"a}ume und eine Klasse von zusammenhangstreuen Abbildungen},
  author={Vietoris, Leopold},
  journal={Mathematische Annalen},
  volume={97},
  number={1},
  pages={454--472},
  year={1927},
  publisher={Springer}
}

@incollection{edelsbrunner2011alpha,
  title={Alpha shapes-a survey},
  author={Edelsbrunner, Herbert},
  booktitle={Tessellations in the sciences: Virtues, techniques and applications of geometric tilings},
  year={2011}
}

@article{kulmanov2018deepgo,
  title={DeepGO: predicting protein functions from sequence and interactions using a deep ontology-aware classifier},
  author={Kulmanov, Maxat and Khan, Mohammed Asif and Hoehndorf, Robert},
  journal={Bioinformatics},
  volume={34},
  number={4},
  pages={660--668},
  year={2018},
  publisher={Oxford University Press}
}

@article{kulmanov2020deepgoplus,
  title={DeepGOPlus: improved protein function prediction from sequence},
  author={Kulmanov, Maxat and Hoehndorf, Robert},
  journal={Bioinformatics},
  volume={36},
  number={2},
  pages={422--429},
  year={2020},
  publisher={Oxford University Press}
}

@article{wang2023mmsmaplus,
  title={MMSMAPlus: a multi-view multi-scale multi-attention embedding model for protein function prediction},
  author={Wang, Zhongyu and Deng, Zhaohong and Zhang, Wei and Lou, Qiongdan and Choi, Kup-Sze and Wei, Zhisheng and Wang, Lei and Wu, Jing},
  journal={Briefings in Bioinformatics},
  volume={24},
  number={4},
  pages={bbad201},
  year={2023},
  publisher={Oxford University Press}
}

@article{song2024deepss2go,
  title={DeepSS2GO: protein function prediction from secondary structure},
  author={Song, Fu V and Su, Jiaqi and Huang, Sixing and Zhang, Neng and Li, Kaiyue and Ni, Ming and Liao, Maofu},
  journal={Briefings in Bioinformatics},
  volume={25},
  number={3},
  pages={bbae196},
  year={2024},
  publisher={Oxford University Press}
}

@article{cao2021tale,
  title={TALE: Transformer-based protein function Annotation with joint sequence--Label Embedding},
  author={Cao, Yue and Shen, Yang},
  journal={Bioinformatics},
  volume={37},
  number={18},
  pages={2825--2833},
  year={2021},
  publisher={Oxford University Press}
}

@article{yang2026multi,
  title={Multi-View Collaboration Feature Fusion for Protein Function Prediction},
  author={Yang, Hailong and Wang, Zhongyu and Shi, Haijun and Ning, Qiao and Deng, Zhaohong and Hu, Shudong and Zhong, Yanqi},
  journal={Journal of Chemical Information and Modeling},
  year={2026},
  publisher={ACS Publications}
}

@article{buchfink2015fast,
  title={Fast and sensitive protein alignment using DIAMOND},
  author={Buchfink, Benjamin and Xie, Chao and Huson, Daniel H},
  journal={Nature methods},
  volume={12},
  number={1},
  pages={59--60},
  year={2015},
  publisher={Nature Publishing Group US New York}
}

@article{radivojac2013large,
  title={A large-scale evaluation of computational protein function prediction},
  author={Radivojac, Predrag and Clark, Wyatt T and Oron, Tal Ronnen and Schnoes, Alexandra M and Wittkop, Tobias and Sokolov, Artem and Graim, Kiley and Funk, Christopher and Verspoor, Karin and Ben-Hur, Asa and others},
  journal={Nature methods},
  volume={10},
  number={3},
  pages={221--227},
  year={2013},
  publisher={Nature Publishing Group US New York}
}

@article{zhu2022integrating,
  title={Integrating unsupervised language model with triplet neural networks for protein gene ontology prediction},
  author={Zhu, Yi-Heng and Zhang, Chengxin and Yu, Dong-Jun and Zhang, Yang},
  journal={PLOS Computational Biology},
  volume={18},
  number={12},
  pages={e1010793},
  year={2022},
  publisher={Public Library of Science San Francisco, CA USA}
}

@article{you2021deepgraphgo,
  title={DeepGraphGO: graph neural network for large-scale, multispecies protein function prediction},
  author={You, Ronghui and Yao, Shuwei and Mamitsuka, Hiroshi and Zhu, Shanfeng},
  journal={Bioinformatics},
  volume={37},
  number={Supplement\_1},
  pages={i262--i271},
  year={2021},
  publisher={Oxford University Press}
}

@article{wang2025dpfunc,
  title={DPFunc: accurately predicting protein function via deep learning with domain-guided structure information},
  author={Wang, Wenkang and Shuai, Yunyan and Zeng, Min and Fan, Wei and Li, Min},
  journal={Nature communications},
  volume={16},
  number={1},
  pages={70},
  year={2025},
  publisher={Nature Publishing Group UK London}
}

@article{gligorijevic2021structure,
  title={Structure-based protein function prediction using graph convolutional networks},
  author={Gligorijevi{\'c}, Vladimir and Renfrew, P Douglas and Kosciolek, Tomasz and Leman, Julia Koehler and Berenberg, Daniel and Vatanen, Tommi and Chandler, Chris and Taylor, Bryn C and Fisk, Ian M and Vlamakis, Hera and others},
  journal={Nature communications},
  volume={12},
  number={1},
  pages={3168},
  year={2021},
  publisher={Nature Publishing Group UK London}
}

@article{lai2022accurate,
  title={Accurate protein function prediction via graph attention networks with predicted structure information},
  author={Lai, Boqiao and Xu, Jinbo},
  journal={Briefings in Bioinformatics},
  volume={23},
  number={1},
  pages={bbab502},
  year={2022},
  publisher={Oxford University Press}
}

@article{lin2025gobeacon,
  title={Gobeacon: an ensemble model for protein function prediction enhanced by contrastive learning},
  author={Lin, Weining and Miller, David and Gu, Zhonghui and Orengo, Christine},
  journal={Protein Science},
  volume={34},
  number={7},
  pages={e70182},
  year={2025},
  publisher={Wiley Online Library}
}

@article{steinegger2017mmseqs2,
  title={MMseqs2 enables sensitive protein sequence searching for the analysis of massive data sets},
  author={Steinegger, Martin and S{\"o}ding, Johannes},
  journal={Nature biotechnology},
  volume={35},
  number={11},
  pages={1026--1028},
  year={2017},
  publisher={Nature Publishing Group US New York}
}

@article{fu2012cd,
  title={CD-HIT: accelerated for clustering the next-generation sequencing data},
  author={Fu, Limin and Niu, Beifang and Zhu, Zhengwei and Wu, Sitao and Li, Weizhong},
  journal={Bioinformatics},
  volume={28},
  number={23},
  pages={3150--3152},
  year={2012},
  publisher={Oxford University Press}
}

@article{jones2014interproscan,
  title={InterProScan 5: genome-scale protein function classification},
  author={Jones, Philip and Binns, David and Chang, Hsin-Yu and Fraser, Matthew and Li, Weizhong and McAnulla, Craig and McWilliam, Hamish and Maslen, John and Mitchell, Alex and Nuka, Gift and others},
  journal={Bioinformatics},
  volume={30},
  number={9},
  pages={1236--1240},
  year={2014},
  publisher={Oxford University Press}
}

@incollection{boutet2007uniprotkb,
  title={UniProtKB/Swiss-Prot: the manually annotated section of the UniProt KnowledgeBase},
  author={Boutet, Emmanuel and Lieberherr, Damien and Tognolli, Michael and Schneider, Michel and Bairoch, Amos},
  booktitle={Plant bioinformatics: methods and protocols},
  pages={89--112},
  year={2007},
  publisher={Springer}
}

@article{eisenberg2000protein,
  title={Protein function in the post-genomic era},
  author={Eisenberg, David and Marcotte, Edward M and Xenarios, Ioannis and Yeates, Todd O},
  journal={Nature},
  volume={405},
  number={6788},
  pages={823--826},
  year={2000},
  publisher={Nature Publishing Group UK London}
}

@article{cho2016compact,
  title={Compact integration of multi-network topology for functional analysis of genes},
  author={Cho, Hyunghoon and Berger, Bonnie and Peng, Jian},
  journal={Cell systems},
  volume={3},
  number={6},
  pages={540--548},
  year={2016},
  publisher={Elsevier}
}

@article{mostafavi2008genemania,
  title={GeneMANIA: a real-time multiple association network integration algorithm for predicting gene function},
  author={Mostafavi, Sara and Ray, Debajyoti and Warde-Farley, David and Grouios, Chris and Morris, Quaid},
  journal={Genome biology},
  volume={9},
  number={Suppl 1},
  pages={S4},
  year={2008},
  publisher={Springer}
}

@inproceedings{kahanda2017gostruct,
  title={Gostruct 2.0: Automated protein function prediction for annotated proteins},
  author={Kahanda, Indika and Ben-Hur, Asa},
  booktitle={Proceedings of the 8th ACM International Conference on Bioinformatics, Computational Biology, and Health Informatics},
  pages={60--66},
  year={2017}
}

@article{zhou2019cafa,
  title={The CAFA challenge reports improved protein function prediction and new functional annotations for hundreds of genes through experimental screens},
  author={Zhou, Naihui and Jiang, Yuxiang and Bergquist, Timothy R and Lee, Alexandra J and Kacsoh, Balint Z and Crocker, Alex W and Lewis, Kimberley A and Georghiou, George and Nguyen, Huy N and Hamid, Md Nafiz and others},
  journal={Genome biology},
  volume={20},
  number={1},
  pages={244},
  year={2019},
  publisher={Springer}
}

@article{jiang2016expanded,
  title={An expanded evaluation of protein function prediction methods shows an improvement in accuracy},
  author={Jiang, Yuxiang and Oron, Tal Ronnen and Clark, Wyatt T and Bankapur, Asma R and D’Andrea, Daniel and Lepore, Rosalba and Funk, Christopher S and Kahanda, Indika and Verspoor, Karin M and Ben-Hur, Asa and others},
  journal={Genome biology},
  volume={17},
  number={1},
  pages={184},
  year={2016},
  publisher={Springer}
}

@article{gene2019gene,
  title={The gene ontology resource: 20 years and still GOing strong},
  author={Gene Ontology Consortium},
  journal={Nucleic acids research},
  volume={47},
  number={D1},
  pages={D330--D338},
  year={2019},
  publisher={Oxford University Press}
}

@article{edelsbrunner2002topological,
  title={Topological persistence and simplification},
  author={Edelsbrunner and Letscher and Zomorodian},
  journal={Discrete \& computational geometry},
  volume={28},
  number={4},
  pages={511--533},
  year={2002},
  publisher={Springer}
}

@article{hozumi2024revealing,
  title={Revealing the shape of genome space via k-mer topology},
  author={Hozumi, Yuta and Wei, Guo-Wei},
  journal={arXiv preprint arXiv:2412.20202},
  year={2024}
}

@article{suwayyid2025cakl,
  title={Cakl: Commutative algebra k-mer learning of genomics},
  author={Suwayyid, Faisal and Hozumi, Yuta and Feng, Hongsong and Zia, Mushal and Wee, JunJie and Wei, Guo-Wei},
  journal={ArXiv},
  pages={arXiv--2508},
  year={2025}
}

@article{grigor2012homologies,
  title={Homologies of path complexes and digraphs},
  author={Grigor'yan, Alexander and Lin, Yong and Muranov, Yuri and Yau, Shing-Tung},
  journal={arXiv preprint arXiv:1207.2834},
  year={2012}
}

@article{grigor2020path,
  title={Path complexes and their homologies},
  author={Grigor’yan, Alexander A and Lin, Yong and Muranov, Yu V and Yau, Shing-Tung},
  journal={Journal of Mathematical Sciences},
  volume={248},
  number={5},
  pages={564--599},
  year={2020},
  publisher={Springer}
}

@inproceedings{li2024path,
  title={Path complex neural network for molecular property prediction},
  author={Li, Longlong and Liu,  Xiang and Wang, Guanghui and Wang, Yu Guang and Xia, Kelin},
  booktitle={ICML 2024 Workshop on Geometry-grounded Representation Learning and Generative Modeling},
  year={2024}
}

@article{thumuluri2022deeploc,
  title={DeepLoc 2.0: multi-label subcellular localization prediction using protein language models},
  author={Thumuluri, Vineet and Almagro Armenteros, Jos{\'e} Juan and Johansen, Alexander Rosenberg and Nielsen, Henrik and Winther, Ole},
  journal={Nucleic acids research},
  volume={50},
  number={W1},
  pages={W228--W234},
  year={2022},
  publisher={Oxford University Press}
}

@article{park2011protein,
  title={Protein localization as a principal feature of the etiology and comorbidity of genetic diseases},
  author={Park, Solip and Yang, Jae-Seong and Shin, Young-Eun and Park, Juyong and Jang, Sung Key and Kim, Sanguk},
  journal={Molecular systems biology},
  volume={7},
  number={1},
  pages={MSB201129},
  year={2011},
  publisher={Springer}
}

@article{briesemeister2010going,
  title={Going from where to why—interpretable prediction of protein subcellular localization},
  author={Briesemeister, Sebastian and Rahnenf{\"u}hrer, J{\"o}rg and Kohlbacher, Oliver},
  journal={Bioinformatics},
  volume={26},
  number={9},
  pages={1232--1238},
  year={2010},
  publisher={Oxford University Press}
}

@article{almagro2017deeploc,
  title={DeepLoc: prediction of protein subcellular localization using deep learning},
  author={Almagro Armenteros, Jos{\'e} Juan and S{\o}nderby, Casper Kaae and S{\o}nderby, S{\o}ren Kaae and Nielsen, Henrik and Winther, Ole},
  journal={Bioinformatics},
  volume={33},
  number={21},
  pages={3387--3395},
  year={2017},
  publisher={Oxford University Press}
}

@article{wan2017fuel,
  title={FUEL-mLoc: feature-unified prediction and explanation of multi-localization of cellular proteins in multiple organisms},
  author={Wan, Shibiao and Mak, Man-Wai and Kung, Sun-Yuan},
  journal={Bioinformatics},
  volume={33},
  number={5},
  pages={749--750},
  year={2017},
  publisher={Oxford University Press}
}

@article{stark2021light,
  title={Light attention predicts protein location from the language of life},
  author={St{\"a}rk, Hannes and Dallago, Christian and Heinzinger, Michael and Rost, Burkhard},
  journal={Bioinformatics Advances},
  volume={1},
  number={1},
  pages={vbab035},
  year={2021},
  publisher={Oxford University Press}
}

@article{meulemans2010defining,
  title={Defining the pathogenesis of the human Atp12p W94R mutation using a Saccharomyces cerevisiae yeast model},
  author={Meulemans, Ann and Seneca, Sara and Pribyl, Thomas and Smet, Joel and Alderweirldt, Valerie and Waeytens, Anouk and Lissens, Willy and Van Coster, Rudy and De Meirleir, Linda and Di Rago, Jean-Paul and others},
  journal={Journal of Biological Chemistry},
  volume={285},
  number={6},
  pages={4099--4109},
  year={2010},
  publisher={Elsevier}
}

@article{rives2021biological,
  title={Biological structure and function emerge from scaling unsupervised learning to 250 million protein sequences},
  author={Rives, Alexander and Meier, Joshua and Sercu, Tom and Goyal, Siddharth and Lin, Zeming and Liu, Jason and Guo, Demi and Ott, Myle and Zitnick, C Lawrence and Ma, Jerry and others},
  journal={Proceedings of the national academy of sciences},
  volume={118},
  number={15},
  pages={e2016239118},
  year={2021},
  publisher={National Academy of Sciences}
}

@article{ravikant2010pie,
  title={Pie—efficient filters and coarse grained potentials for unbound protein--protein docking},
  author={Ravikant, DVS and Elber, Ron},
  journal={Proteins: Structure, Function, and Bioinformatics},
  volume={78},
  number={2},
  pages={400--419},
  year={2010},
  publisher={Wiley Online Library}
}

@article{abbasi2020island,
  title={ISLAND: in-silico proteins binding affinity prediction using sequence information},
  author={Abbasi, Wajid Arshad and Yaseen, Adiba and Hassan, Fahad Ul and Andleeb, Saiqa and Minhas, Fayyaz Ul Amir Afsar},
  journal={BioData Mining},
  volume={13},
  number={1},
  pages={20},
  year={2020},
  publisher={Springer}
}

@article{romero2022ppi,
  title={PPI-affinity: A web tool for the prediction and optimization of protein--peptide and protein--protein binding affinity},
  author={Romero-Molina, Sandra and Ruiz-Blanco, Yasser B and Mieres-Perez, Joel and Harms, Mirja and Munch, Jan and Ehrmann, Michael and Sanchez-Garcia, Elsa},
  journal={Journal of proteome research},
  volume={21},
  number={8},
  pages={1829--1841},
  year={2022},
  publisher={ACS Publications}
}

@article{zhou2024proaffinity,
  title={ProAffinity-GNN: a novel approach to structure-based protein--Protein binding affinity prediction via a curated data set and graph neural networks},
  author={Zhou, Zhiyuan and Yin, Yueming and Han, Hao and Jia, Yiping and Koh, Jun Hong and Kong, Adams Wai-Kin and Mu, Yuguang},
  journal={Journal of chemical information and modeling},
  volume={64},
  number={23},
  pages={8796--8808},
  year={2024},
  publisher={ACS Publications}
}

@article{xie2025ppi,
  title={PPI-Graphomer: enhanced protein-protein affinity prediction using pretrained and graph transformer models},
  author={Xie, Jun and Zhang, Youli and Wang, Ziyang and Jin, Xiaocheng and Lu, Xiaoli and Ge, Shengxiang and Min, Xiaoping},
  journal={BMC bioinformatics},
  volume={26},
  number={1},
  pages={116},
  year={2025},
  publisher={Springer}
}

@article{xu2025ssif,
  title={SSIF-Affinity: Multimodal Deep Learning of Sequence-Structure Features for Precise Protein--Protein Binding Affinity Prediction},
  author={Xu, Xinyi and Zhang, Haotian and Liu, Qi and Cheng, Qian and Guo, Yanbei and Yang, Wei and Chen, Xueli},
  journal={Journal of Chemical Information and Modeling},
  volume={66},
  number={1},
  pages={74--87},
  year={2025},
  publisher={ACS Publications}
}

@article{vangone2015contacts,
  title={Contacts-based prediction of binding affinity in protein--protein complexes},
  author={Vangone, Anna and Bonvin, Alexandre MJJ},
  journal={elife},
  volume={4},
  pages={e07454},
  year={2015},
  publisher={eLife Sciences Publications, Ltd}
}

@article{yang2021pon,
  title={PON-Sol2: prediction of effects of variants on protein solubility},
  author={Yang, Yang and Zeng, Lianjie and Vihinen, Mauno},
  journal={International Journal of Molecular Sciences},
  volume={22},
  number={15},
  pages={8027},
  year={2021},
  publisher={MDPI}
}

@article{paladin2017soda,
  title={SODA: prediction of protein solubility from disorder and aggregation propensity},
  author={Paladin, Lisanna and Piovesan, Damiano and Tosatto, Silvio CE},
  journal={Nucleic acids research},
  volume={45},
  number={W1},
  pages={W236--W240},
  year={2017},
  publisher={Oxford University Press}
}

@article{yang2016pon,
  title={PON-Sol: prediction of effects of amino acid substitutions on protein solubility},
  author={Yang, Yang and Niroula, Abhishek and Shen, Bairong and Vihinen, Mauno},
  journal={Bioinformatics},
  volume={32},
  number={13},
  pages={2032--2034},
  year={2016},
  publisher={Oxford University Press}
}

@article{wee2024integration,
  title={Integration of persistent Laplacian and pre-trained transformer for protein solubility changes upon mutation},
  author={Wee, JunJie and Chen, Jiahui and Xia, Kelin and Wei, Guo-Wei},
  journal={Computers in biology and medicine},
  volume={169},
  pages={107918},
  year={2024},
  publisher={Elsevier}
}

@article{huntley2015goa,
  title={The GOA database: gene ontology annotation updates for 2015},
  author={Huntley, Rachael P and Sawford, Tony and Mutowo-Meullenet, Prudence and Shypitsyna, Aleksandra and Bonilla, Carlos and Martin, Maria J and O'Donovan, Claire},
  journal={Nucleic acids research},
  volume={43},
  number={D1},
  pages={D1057--D1063},
  year={2015},
  publisher={Oxford University Press}
}

@article{edgar2010search,
  title={Search and clustering orders of magnitude faster than BLAST},
  author={Edgar, Robert C},
  journal={Bioinformatics},
  volume={26},
  number={19},
  pages={2460--2461},
  year={2010},
  publisher={Oxford University Press}
}

@article{kastritis2011structure,
  title={A structure-based benchmark for protein--protein binding affinity},
  author={Kastritis, Panagiotis L and Moal, Iain H and Hwang, Howook and Weng, Zhiping and Bates, Paul A and Bonvin, Alexandre MJJ and Janin, Jo{\"e}l},
  journal={Protein Science},
  volume={20},
  number={3},
  pages={482--491},
  year={2011},
  publisher={Wiley Online Library}
}

@misc{esm2024cambrian,
  author = {{ESM Team}},
  title = {ESM Cambrian: Revealing the mysteries of proteins with unsupervised learning},
  year = {2024},
  publisher = {EvolutionaryScale Website},
  url = {https://evolutionaryscale.ai/blog/esm-cambrian},
  urldate = {2024-12-04}
}

@article{guo2008using,
  title={Using support vector machine combined with auto covariance to predict protein--protein interactions from protein sequences},
  author={Guo, Yanzhi and Yu, Lezheng and Wen, Zhining and Li, Menglong},
  journal={Nucleic acids research},
  volume={36},
  number={9},
  pages={3025--3030},
  year={2008},
  publisher={Oxford University Press}
}

@article{ullanat2026learning,
  title={Learning the language of protein-protein interactions},
  author={Ullanat, Varun and Jing, Bowen and Sledzieski, Samuel and Berger, Bonnie},
  journal={Nature Communications},
  year={2026},
  publisher={Nature Publishing Group UK London}
}

@article{elnaggar2021prottrans,
  title={ProtTrans: toward understanding the language of life through self-supervised learning},
  author={Elnaggar, Ahmed and Heinzinger, Michael and Dallago, Christian and Rehawi, Ghalia and Wang, Yu and Jones, Llion and Gibbs, Tom and Feher, Tamas and Angerer, Christoph and Steinegger, Martin and others},
  journal={IEEE transactions on pattern analysis and machine intelligence},
  volume={44},
  number={10},
  pages={7112--7127},
  year={2021},
  publisher={IEEE}
}

@article{nijkamp2023progen2,
  title={Progen2: exploring the boundaries of protein language models},
  author={Nijkamp, Erik and Ruffolo, Jeffrey A and Weinstein, Eli N and Naik, Nikhil and Madani, Ali},
  journal={Cell systems},
  volume={14},
  number={11},
  pages={968--978},
  year={2023},
  publisher={Elsevier}
}

@inproceedings{nguyen2010cosine,
  title={Cosine similarity metric learning for face verification},
  author={Nguyen, Hieu V and Bai, Li},
  booktitle={Asian conference on computer vision},
  pages={709--720},
  year={2010},
  organization={Springer}
}

@article{liu2022dowker,
  title={Dowker complex based machine learning (DCML) models for protein-ligand binding affinity prediction},
  author={Liu, Xiang and Feng, Huitao and Wu, Jie and Xia, Kelin},
  journal={PLoS computational biology},
  volume={18},
  number={4},
  pages={e1009943},
  year={2022},
  publisher={Public Library of Science San Francisco, CA USA}
}

@article{grigor2023homotopy,
  title={Homotopy theory for digraphs},
  author={Grigor’yan, Alexander and Lin, Yong and Muranov, Yuri and Yau, Shing-Tung},
  journal={Pure and Applied Mathematics Quarterly},
  volume={10},
  number={4},
  pages={619--674},
  year={2023},
  publisher={International Press of Boston, Inc. Somerville, MA 02143, USA}
}

@article{jankauskaite2019skempi,
  title={SKEMPI 2.0: an updated benchmark of changes in protein--protein binding energy, kinetics and thermodynamics upon mutation},
  author={Jankauskait{\.e}, Justina and Jim{\'e}nez-Garc{\'\i}a, Brian and Dapk{\=u}nas, Justas and Fern{\'a}ndez-Recio, Juan and Moal, Iain H},
  journal={Bioinformatics},
  volume={35},
  number={3},
  pages={462--469},
  year={2019},
  publisher={Oxford University Press}
}

@article{chithrananda2020chemberta,
  title={ChemBERTa: large-scale self-supervised pretraining for molecular property prediction},
  author={Chithrananda, Seyone and Grand, Gabriel and Ramsundar, Bharath},
  journal={arXiv preprint arXiv:2010.09885},
  year={2020}
}

@article{wang2025multicycpermea,
  title={MultiCycPermea: accurate and interpretable prediction of cyclic peptide permeability using a multimodal image-sequence model},
  author={Wang, Zixu and Chen, Yangyang and Shang, Yifan and Yang, Xiulong and Pan, Wenqiong and Ye, Xiucai and Sakurai, Tetsuya and Zeng, Xiangxiang},
  journal={BMC biology},
  volume={23},
  number={1},
  pages={63},
  year={2025},
  publisher={Springer}
}

@article{cao2024multi_cycgt,
  title={Multi\_CycGT: a deep learning-based multimodal model for predicting the membrane permeability of cyclic peptides},
  author={Cao, Lujing and Xu, Zhenyu and Shang, Tianfeng and Zhang, Chengyun and Wu, Xinyi and Wu, Yejian and Zhai, Silong and Zhan, Zhajun and Duan, Hongliang},
  journal={Journal of medicinal chemistry},
  volume={67},
  number={3},
  pages={1888--1899},
  year={2024},
  publisher={ACS Publications}
}

@article{lu2026mfcpedpred,
  title={MFCPepPred: prediction of multi-functional cyclic peptides using tensor-based multi-modal feature fusion},
  author={Lu, Xun and Zhang, Daijun and Wu, Xiangning and Bin, Yannan},
  journal={submitted},
  year={2026},
}

@article{liu2024cyclicpepedia,
  title={CyclicPepedia: a knowledge base of natural and synthetic cyclic peptides},
  author={Liu, Lei and Yang, Liu and Cao, Suqi and Gao, Zhigang and Yang, Bin and Zhang, Guoqing and Zhu, Ruixin and Wu, Dingfeng},
  journal={Briefings in bioinformatics},
  volume={25},
  number={3},
  pages={bbae190},
  year={2024},
  publisher={Oxford University Press}
}

\newpage
\section*{Supplementary Information}


\setcounter{section}{0}
\section{Summary of Datasets}
Dataset details are crucial for  machine learning prediction, transparency, and reproducibility. For each given dataset, the performance of different machine learning algorithms should only be compared with the same training, validation, and test splitting to ensure fairness. Table \ref{tab:all_datasets} presents a brief summary of all datasets used in this work.   

\begin{table*}[htbp]
\centering
\scriptsize
\caption{Summary of datasets used in this study. MLBC: multi-label binary classification; CV: cross-validation; CC: cellular component; BP: biological process; MF: molecular function.}
\label{tab:all_datasets}

\begin{tblr}{
  width=\textwidth,
  colspec={
    Q[l,m,wd=3.0cm]
    Q[c,m,wd=2.5cm]
    Q[c,m,wd=0.9cm]
    Q[c,m,wd=1.35cm]
    Q[r,m,wd=1.25cm]
    Q[r,m,wd=1.05cm]
    Q[r,m,wd=1.15cm]
    Q[r,m,wd=1.25cm]
  },
  row{1}={font=\bfseries},
  column{1}={bg=gray!8},
  hline{1,Z}={1pt},
  hline{2}={0.7pt},
  hline{5,8,11,14,16,17,18,19,20,21}={0.45pt},
  rowsep=2.2pt,
  colsep=3.5pt,
}
Task category 
& Dataset 
& Subtask 
& Task type 
& Train 
& Val. 
& Test 
& Total \\

\SetCell[r=12]{l,m,bg=gray!8} Protein function prediction
& \SetCell[r=3]{c,m} CAFA3 \cite{zhou2019cafa}
& CC & MLBC & 49,331 & 1,265 & 1,265 & 51,861 \\
& & BP & MLBC & 51,108 & 2,392 & 2,392 & 55,892 \\
& & MF & MLBC & 34,973 & 1,137 & 1,137 & 37,247 \\

& \SetCell[r=3]{c,m} CAFA4 \cite{wang2025dpfunc}
& CC & MLBC & 42,837 & 716 & 897 & 44,450 \\
& & BP & MLBC & 47,774 & 776 & 1,068 & 49,618 \\
& & MF & MLBC & 31,732 & 684 & 407 & 32,823 \\

& \SetCell[r=3]{c,m} PDB \cite{gligorijevic2021structure} 
& CC & MLBC & 11,307 & 1,300 & 3,415 & 16,022 \\
& & BP & MLBC & 23,527 & 2,627 & 3,415 & 29,569 \\
& & MF & MLBC & 24,977 & 2,750 & 3,414 & 31,141 \\

& \SetCell[r=3]{c,m} SwissProt \cite{boutet2007uniprotkb}
& CC & MLBC & \SetCell[c=3]{c} Five-fold CV & & & 466,475 \\
& & BP & MLBC & \SetCell[c=3]{c} Five-fold CV & & & 464,711 \\
& & MF & MLBC & \SetCell[c=3]{c} Five-fold CV & & & 491,528 \\

\SetCell[r=2]{l,m,bg=gray!8} Protein subcellular localization prediction
& SPL+HPA \cite{thumuluri2022deeploc}
& -- & MLBC & 28,303 & -- & 1,717 & 30,020 \\
& SPL \cite{thumuluri2022deeploc}
& -- & MLBC & \SetCell[c=3]{c} Five-fold CV & & & 28,303 \\

\SetCell{l,m,bg=gray!8} Mutation-induced protein solubility prediction
& PON-Sol \cite{yang2021pon}
& -- & 3-class & 5,666 & -- & 662 & 6,328 \\

\SetCell{l,m,bg=gray!8} Protein--protein interaction classification
& Yeast-PPI \cite{guo2008using}
& -- & Binary & 4,945 & 95 & 394 & 5,434 \\

\SetCell{l,m,bg=gray!8} Protein--protein binding affinity prediction
& PDBBind+S79 \cite{xu2025ssif,kastritis2011structure}
& -- & Regression & 2,492 & -- & 79 & 2,571 \\

\SetCell{l,m,bg=gray!8} Mutation-induced protein--protein binding affinity prediction
& SKEMPI \cite{jankauskaite2019skempi}
& -- & Regression & \SetCell[c=3]{c} Three-fold CV & & & 6,706 \\

\SetCell{l,m,bg=gray!8} Cyclic peptide function prediction
& CyclicPepedia \cite{lu2026mfcpedpred}
& -- & MLBC & \SetCell[c=3]{c} Five-fold CV & & & 1,322 \\
\end{tblr}
\end{table*}

\section{Mathematical Theories}
\subsection{Category Theory of Persistent Laplacians}
The original topological spectral theory of persistent Laplacian was introduced by Wang et al. in 2019 \cite{wang2020persistent}. 
Here, we provide the functorial construction of the persistent Laplacian in terms of category theory. 
\subsubsection{Category and Functor}
\begin{definition}
    A category $\mathscr{C}$ consists of
    \begin{itemize}
        \item a collection of objects, denoted by $ob(\mathscr{C})$,
        \item for each pair of objects $A,B\in ob(\mathscr{C})$, there is a morphism set $\mathscr{C}(A,B)$ consisting of maps from $A$ to $B$,
        \item for each $A,B,C\in ob(\mathscr{C})$, there is a composition of $\mathscr{C}(A,B)$ and $\mathscr{C}(B,C)$ to $\mathscr{C}(A,C)$,
        \item for each object $A\in\mathscr{C}$, there is an identity morphism $1_A\in\mathscr{C}(A,A)$.
    \end{itemize}
    satisfying the following commutative diagrams for $\forall$ $f\in\mathscr{C}(A,B)$, $g\in\mathscr{C}(B,C)$, $h\in\mathscr{C}(C,D)$:

    \begin{equation}\nonumber
         \begin{tikzcd}
            A \arrow[r, "f"] \arrow[d, "gf"'] & B \arrow[d, "hg"] &  & A \arrow[r, "1_A"] \arrow[d, "f"'] \arrow[rd, "f"] & A \arrow[d, "f"] \\
            C \arrow[r, "h"]                  & D                 &  & B \arrow[r, "1_B"']                                & B               
        \end{tikzcd}
    \end{equation}

\end{definition}

\begin{example}
    The category $(\mathbb{R},\leq)$ is the category whose objects are real numbers and whose morphisms are given by the order relation $a\leq b$. Similarly, $\mathbf{Set}$ denotes the category whose objects are sets and whose morphisms are maps between sets. When the composition law and identity morphisms are canonical or clear from context, we often omit their explicit specification and simply refer to $\mathbf{Set}$ as the category of sets. For example, the category of simplicial complexes is the category whose objects are simplicial complexes and whose morphisms are simplicial maps.
\end{example}

\begin{definition}
    Given two categories $\mathscr{A}$ and $\mathscr{B}$, a functor F from $\mathscr{A}$ to $\mathscr{B}$ consists of 
    \begin{itemize}
        \item a function F: $ob(\mathscr{A})\to ob(\mathscr{B})$ as $A\mapsto \text{F}(A)$,
        \item a function F: $\mathscr{A}(A,A')\to\mathscr{B}(\text{F}(A),\textbf{F}(A'))$ as $f\mapsto \textbf{F}(f)$ for all $A,A'\in ob(\mathscr{A})$ and $f\in \mathscr{A}(A,A')$.
    \end{itemize}
    satisfying the following commutative diagrams:
    \begin{equation}\nonumber
        \begin{tikzcd}
        A \arrow[r, "f"] \arrow[d, "F"] & B \arrow[r, "g"]       & C \arrow[d, "F"] &  & A \arrow[r, "F"] \arrow[d, "1_A"] & F(A) \arrow[d, "1_{F(A)}"] \\
        F(A) \arrow[r, "F(f)"]          & F(B) \arrow[r, "F(g)"] & F(C)             &  & A \arrow[r, "F"]                  & F(A)                      
        \end{tikzcd}
    \end{equation}
    
\end{definition}

\begin{example}
    The notation of functor establishes connections between different categories. For example, the map from the category of simplicial complexes to the category of sets is a functor. The homology group of simplicial complexes is a functor from the category of simplicial complexes to the category of groups. 
\end{example}

\begin{definition}
    Given a category $\mathscr{C}$, a persistence object on $\mathscr{C}$ is a functor $\mathcal{F}:(\mathbb{R},\leq)\to \mathscr{C}$ from the category $(\mathbb{R},\leq)$ to the category $\mathscr{C}$.
\end{definition}

\begin{example}
    Let \textbf{Simp} be the category of simplicial complexes, then the functor $\mathcal{F}:(\mathbb{R},\leq)\to $ \textbf{Simp} is a persistence simplicial complex or a filtration of simplicial complexes. The commonly used Vietoris-Rips complex and Alpha complex are persistence simplicial complexes.
\end{example}

\subsubsection{Differential Graded Inner Product Space}
\begin{definition}
    A differential graded inner product space $V=\{V_i\}$ is a graded algebraic object equipped with a differential $d$ satisfying $d^2=0$ and an inner product $\langle\cdot,\cdot\rangle$, which allows one to define adjoint operators.
    \begin{equation}\nonumber
        \begin{tikzcd}
        ... \arrow[r] & V_{k+1} \arrow[r, "d"] & V_k \arrow[r, "d", bend left] & V_{k-1} \arrow[r, "d"] \arrow[l, "d^*", bend left] & ...
        \end{tikzcd}
    \end{equation}
\end{definition}

Throughout this section, all differential graded inner product spaces are assumed to be finite-dimensional.
Given a differential graded inner product space $V$ and a linear operator $T:V\to V$, we denote by $T^*$ the adjoint of $T$. We have
$$T^{**}=T,\qquad \langle x,Ty\rangle=\langle T^*x,y\rangle$$
for all $x,y\in V$.
\begin{definition}
    A morphism $f:V\to W$ of differential graded inner product spaces is a chain map that satisfies the following commutative diagrams:

    \begin{equation}\nonumber
         \begin{tikzcd}
        ... \arrow[r] & V_{k+1} \arrow[r, "d^V"] \arrow[d, "f"] & V_k \arrow[r, "d^V"] \arrow[d, "f"] & V_{k-1} \arrow[r] \arrow[d, "f"] & ... \\
        ... \arrow[r] & W_{k+1} \arrow[r, "d^W"']               & W_k \arrow[r, "d^W"']               & W_{k-1} \arrow[r]                & ...
        \end{tikzcd}
    \end{equation}

    and preserves the inner product, namely
$$\langle x,y\rangle_V=\langle f(x),f(y)\rangle_W.$$
\end{definition}
Consequently, every morphism of differential graded inner product spaces is injective, and
$$f^*f=\mathrm{id}_V.$$
\begin{remark}
    Let $\mathbf{DGI}$ denote the category of finite-dimensional differential graded inner product spaces over $\mathbb{R}$. Throughout the remainder of this work, we use $d$ and $\langle\cdot,\cdot\rangle$ to denote the differentials and inner products of the differential graded inner product spaces under consideration.
\end{remark}

\begin{example}
    Let $\mathbf{Inn}$ be the category of inner product spaces. We have two functors from the category $\mathbf{DGI}$ of differential graded inner product spaces to the category $\mathbf{Inn}$ of inner product spaces: the homology functor $H:\mathbf{DGI}\to\mathbf{Inn}$ and the harmonic space functor $h:\mathbf{DGI}\to \mathbf{Inn}$.  Actually, there is a natural isomorphism $\rho:h\to H$ of these two functors.
\end{example}

\subsubsection{Laplacian}
Given a morphism $f:V\to W$ of differential graded inner product spaces, define the subspace $\Theta_f$ of $W$ by
$$\Theta_f=\{x\in W|dx\in f(V)\}.$$
Then we have a short sequence
$$\Theta_f\xrightarrow{f^*di}V\xrightarrow{d}V\to0,$$
where $i:\Theta_f\hookrightarrow W$ denotes the inclusion map. For any $x\in \Theta_f$, we have $d(x)=f(v)$ for some $v\in V$. It follows that
$$fdf^*di=fdf^*f(v)=fd(v)=df(v)=d^2x=0$$
Since $f$ is injective, we have $df^*di(x)=0$. Consequently, we have $df^*di=0$. 
\begin{remark}
    Note that all spaces are graded. The $p$-th component of this construction is illustrated by the following diagram:
    \begin{equation}\nonumber
        \begin{tikzcd}[column sep=huge]
        {\Theta_{f,p+1}} \arrow[r, "(f_p)^*d_{p+1}i_{p+1}"] \arrow[d, "i_{p+1}"] & V_p \arrow[r, "d_p"] \arrow[d, "f_p"] & V_{p-1} \\
        W_{p+1} \arrow[r, "d_{p+1}"] & W_p      &        
        \end{tikzcd}
    \end{equation}
\end{remark}
\begin{definition}
    The Laplacian associated with the morphism $f:V\to W$ in dimension $p$ is defined by
$$\Delta_{f,p}=(f_p)^*d_{p+1}i_{p+1}(i_{p+1})^*(d_{p+1})^*f_p+(d_p)^*d_p$$
This operator is self-adjoint and semi-positive definite.
\end{definition}
using this language, the standard persistent laplacian can be defined as follows
\begin{definition}
    For any real numbers $a\leq b$, the $(a,b)$-persistent Laplacian for a persistence differential graded inner product space $\mathcal{F}:(\mathbb{R},\leq)\to \textbf{DGI}$ is defined by
    $$\Delta_\mathcal{F}^{a,b}=(f_{a,b})^*dii^*d^*f_{a,b}+d^*d$$
    where $f_{a,b}:\mathcal{F}_a\to\mathcal{F}_b$ is the morphism of differential graded inner product spaces.
\end{definition}
A key property of the persistent Laplacian is that its kernel is isomorphic to the corresponding persistent homology group of the differential graded inner product space.
\begin{theorem}
    Given a persistence differential graded inner product space $\mathcal{F}:(\mathbb{R},\leq)\to\mathbf{DGI}$. For any $a\leq b$, we have
    $$\text{ker}\Delta_\mathcal{F}^{a,b}\cong H^{a,b}(\mathcal{F})$$
    where $H^{a,b}(\mathcal{F})$ is the $(a,b)$-persistent homology of $\mathcal{F}$.
\end{theorem}

\subsubsection{Category Lap of Laplacian Trees}
\begin{definition}
    Let $V$ be a differential graded inner product space, and let $M(V)$ denote the set of all morphisms of the form $f:V\to W$ in $\mathbf{DGI}$. The Laplacian space $L(V)$ of $V$ is defined to be the set of all Laplacians $\Delta_f$ induced by morphisms $f:V\to W$ in $M(V)$.
\end{definition}
Equivalently, $L(V)$ can be reviewed as the quotient of $M(V)$ by the equivalence relation 
$$f\sim g ~~\text{if}~~ \Delta_f=\Delta_g$$
Given a morphism $\phi:V\to X$ in $\mathbf{DGI}$, it induces a map $L(\phi):L(V)\to L(X)$ of Laplacian spaces, 
$$L(\phi):L(V)\to L(X)$$
defined by
$$L(\phi)(\Delta_f)=\Delta_g,$$
where $g$ is obtained from the following pushout diagram:
\begin{equation}
    \begin{tikzcd}
    V \arrow[r, "f"] \arrow[d, "\phi"] & W \arrow[d, "\psi"] \\
    X \arrow[r, "g"]                   & Y                  
    \end{tikzcd}
\end{equation}
With this well-defined morphism of Laplacian spaces, we get the category of Laplacian trees. 
\begin{definition}
    The category $\mathbf{Lap}$ of Laplacian trees is defined as follows:
    \begin{itemize}
        \item objects are pairs $(V,A)$, where $V$ is a differential graded inner product space in $\mathbf{DGI}$ and $A\subseteq L(V)$ is a subset of its Laplacian space,
        \item morphisms are pairs
        \[
        (\phi,\Phi):(V,A)\to(W,B),
        \]
        where $\phi:V\to W$ is a morphism in $\mathbf{DGI}$ and
        \[
        \Phi=L(\phi)|_A:A\to B
        \]
        is the restriction of the induced map $L(\phi)$ to $A$. 
    \end{itemize}
\end{definition}
Moreover, one can verify that $\mathcal{L}:\textbf{DGI}\to \textbf{Lap}$ 
$$V\mapsto(V,L(V))$$
defines a functor.

\subsubsection{Persistent Laplacian Functor}
Consider a persistence differential graded inner product space 
$$\mathcal{S}:(\mathbb{R},\leq)\to\textbf{DGI}.$$
For each $a\in\mathbb{R}$, define 
$$\mathcal{L}_\mathcal{S}(a)=(\mathcal{S}_a,L^a),$$ 
where $L^a=\{\Delta^{a,t}\}_{a\leq t}$. For $a\leq b$, let $f_{a,b}:\mathcal{S}_a\to \mathcal{S}_b$
be the structure morphism of the persistence object.
we have the morphisms
$$\mathcal{L}_{\mathcal{S}}^{a,b}=(f_{a,b},L(f_{a,b})|_{L^a}):\mathcal{L}_\mathcal{S}(a)\to \mathcal{L}_\mathcal{S}(b), ~~(\mathcal{S}_a,L^a)\mapsto (\mathcal{S}_b,L^b)$$
where the induced map on Laplacian spaces is given by
$$L(f_{a,b})|_{L^a}(\Delta^{a,t})=\Delta^{b,\text{max}\{b,t\}},$$ 
for $\Delta^{a,t}\in L^a$. For any $a\leq b\leq c$, we have
$$L(f_{b,c})L(f_{a,b})(\Delta^{a,t})=L(f_{b,c})(\Delta^{b,\text{max}\{b,t\}})=\Delta^{c,\text{max}\{c,t\}}=L(f_{a,c}(\Delta^{a,t}))$$
Thus the composition law is satisfied. The identity condition follows directly from
$f_{a,a}=\mathrm{id}_{\mathcal{S}_a}.$ Therefore, we obtain a functor
$$\mathcal{L}_\mathcal{S}:(\mathbb{R},\leq)\to\textbf{Lap},~~a\mapsto(\mathcal{S}_a,L^a)$$
\begin{remark}
    Given a persistent simplicial complex, its induced persistent chain complex equipped with the standard inner product defines a persistent differential graded inner product space.
    $$\mathcal{S}:(\mathbb{R},\leq)\to \mathbf{DGI}.$$
    For each filtration value $a$, the object
    $$\mathcal{L}_{\mathcal{S}}(a)=(\mathcal{S}_a,L^a)$$
    consists of the chain complex $\mathcal{S}_a$ at filtration value $a$ together with the set $L^a$ of all persistent Laplacians starting from $a$. Consequently, the persistent Laplacian associated with a persistent simplicial complex can be formulated as a functor from the category $(\mathbb{R},\leq)$ to the category $\mathbf{Lap}$.
\end{remark}

\subsection{Path Complex}
We present a brief summary of the relevant aspects of path complex,  which was introduced by Grigor'yan et al.\cite{grigor2012homologies} 
\begin{definition}
    Given a nonempty finite set G, for any nonnegative integer $p$, an elementary $p$-path on G is any sequence $g_0g_1\cdots g_p$ of $p+1$ vertices of V. For $p=-1$, any elementary $p$-path is the empty set $\emptyset$.
\end{definition}
Denote by $G_k$ the set of all elementary $k$-paths on $G$, $G_{-1}$ is the empty set. Let $\Lambda_k(G)$ be the vector space spanned by $G_k$, we have the standard boundary operator 
\begin{equation}
    \partial:\Lambda_k(G)\to \Lambda_{k-1}(G)
\end{equation}
For any $k$-path $\sigma_k=g_0g_1\cdots g_k$,
\begin{equation}
    \partial(\sigma_k)=\sum_{i=0}^k(-1)^ig_0\cdots g_{i-1}g_{i+1}\cdots g_k
\end{equation}
and this boundary operator has the property that $\partial\partial=0$. Consequently, we get a chain complex 
\begin{equation}\label{eq:chain-complex}
    \cdots\to\Lambda_k(G)\xrightarrow{\partial_k}\Lambda_{k-1}(G)\xrightarrow{\partial_{k-1}}\Lambda_{k-2}(G)\to\cdots\to\Lambda_1(G)\xrightarrow{\partial_1}\Lambda_0(G)\to0.
\end{equation}

\begin{definition}
    A path complex $\mathcal{P}$ over a set G is a nonempty collection of elementary paths on G  satisfies the following condition
    $$\forall g_0g_1\cdots g_n\in\mathcal{P},~g_1\cdots g_n\in\mathcal{P},~g_0g_1\cdots g_{n-1}\in\mathcal{P} $$
\end{definition}
Path complexes generalize the notions of simplicial complexes and directed graphs. An oriented simplicial complex can be naturally regarded as a path complex by treating the oriented simplices as paths. A key property is that one can construct a chain complex for path complex and hence define homology groups that characterize the topological information encoded by the path complex. We now briefly describe the construction of the homology theory for path complexes.

As shown in formula \ref{eq:chain-complex}, we already have a china complex from the vertex set G. However, this chain complex depends only on the underlying vertex set $G$, and therefore does not capture the specific structure of the path complex $\mathcal{P}$. To encode the information of $\mathcal{P}$ itself, one needs to construct a chain complex adapted to the paths in $\mathcal{P}$.

Let $\mathcal{P}_k$ be the set of all $k$-paths in $\mathcal{P}$, and $\Omega_k(\mathcal{P})$ be the vector space spanned by $\mathcal{P}_k$. If we directly restrict the boundary operator $\partial$ to the space $\mathcal{A}_k(\mathcal{P})$, the resulting map may not be well defined as a boundary operator on $\mathcal{A}_*(\mathcal{P})$, because in general $\partial(\mathcal{A}_k(\mathcal{P}))\not\subset\mathcal{A}_{k-1}(\mathcal{P})$. Instead, we consider the subspace 
$$\Omega_k(\mathcal{P})=\{\alpha\in\mathcal{A}_k(\mathcal{P})|\partial(\alpha)\in\mathcal{A}_{k-1}(\mathcal{P})\}$$
This construction will make sure that $\partial_k(\mathcal{A}_k(\mathcal{P}))\subset\mathcal{A}_{k-1}(\mathcal{P})$. Therefore, we obtain a chain complex
\begin{equation}
    \cdots\to\Omega_{k}(\mathcal{P})\xrightarrow{\partial_k}\Omega_{k-1}(\mathcal{P})\xrightarrow{\partial_{k-1}}\Omega_{k-2}(\mathcal{P})\to\cdots\to\Omega_1(\mathcal{P})\xrightarrow{\partial_1}\Omega_0(\mathcal{P})\to0
\end{equation}

The $k$-th homology of the chain complex is defined as the $k$-th homology of the path complex $\mathcal{P}$. As in simplicial homology, the rank of the homology is defined as the Betti number.
These Betti numbers characterize topological features of the path complex, such as connected components, loops, voids, and their higher-dimensional analogues.

\section{Model Implementation Details }
\subsection{Hyperparameter Setting}
In our implementation, persistent Laplacian embeddings are computed for each protein sequence using multiple sliding-window sizes, namely 4, 8, 12, 16, 20, 25, and 30, thereby enabling the characterization of subsequence patterns at different scales. The sliding step is set equal to the corresponding window size. For persistent Laplacian construction, cosine distance is adopted, whose range is [0,2]. The filtration goes from 0 to 0.6 with a step of 0.02, and from 0.6 to 2 with a step of 0.05, yielding a total of 40 filtration values. At each filtration step, we extract the harmonic spectrum of $L_0$, namely the number of zero eigenvalues, together with six statistical descriptors of the non-harmonic spectrum, including the maximum, minimum, mean, standard deviation, sum, and number of nonzero eigenvalues.

For the path complex representation, we adopt a simple grouping strategy in which the 20 amino acid types are treated as 20 distinct groups. A 2-dimensional path complex is employed, consisting of 0-paths, 1-paths, and 2-paths. For each path, the radial basis function (RBF), defined as $rbf(w)=\exp{\left(\frac{-(w-c)^2}{\sigma}\right)}$, is applied to the path weight $w$ to generate a 192-dimensional feature vector using 64 centers $c$ and three variance parameters $\sigma$, where $c\in[0,1]$ with step size 1/64, and $\sigma\in\{1,0.1,0.01\}$. This vector is then concatenated with the ProtTrans embedding to form the initial feature representation of each path. For 1-paths and 2-paths, the average ProtTrans embedding of the amino acids involved in the path is used.

For the neural network component, we employ a three-layer architecture. In each layer, the number of attention heads is set to 4 and the hidden dimension is 256. Model optimization is performed using the AdamW optimizer with a weight decay of 1e-3, together with a one-cycle learning rate scheduler. For the protein function prediction, cyclic peptide function prediction, and protein subcellular localization tasks, Binary Cross-Entropy with Logits is used as the loss function. For the mutation-induced protein solubility prediction task, Cross-Entropy loss is adopted. For the protein--protein binding affinity prediction task, Smooth L1 loss is employed. For the protein--protein binding binary classification task, Cross-Entropy loss is used. For the mutation-induced protein--protein binding affinity changes prediction, the MSE loss is employed.

The model is implemented using PyTorch, and all experiments are conducted on an NVIDIA Tesla V100S 32GB GPU. For most tasks, the model is trained for 10 epochs. The only exceptions are cyclic peptide function prediction and solubility prediction, for which the models are trained for 40 and 20 epochs, respectively. For five-fold cross-validation on the SwissProt dataset, the learning rate and batch size are set to 1e-4 and 128, respectively. For protein function prediction, the learning rates are 5e-5, 2e-4, and 5e-4 for the CAFA3, CAFA4, and PDB datasets, respectively, with a batch size of 32 for all three datasets. The model achieving the best Fmax+AUPR on the validation set is selected as the final model for these datasets. For protein subcellular localization prediction, a learning rate of 5e-5 and batch size of 64 are used. For mutation-induced protein solubility prediction, the learning rate is 3e-4 with a batch size of 32. For protein--protein binding affinity prediction, the learning rate is 2e-5 with batch size 64. For protein--protein interaction binary classification, the learning rate is 1e-4 with batch size 64. For mutation-induced protein--protein binding affinity changes prediction, the learning rate is 5e-4 with batch size 32. For cyclic peptide function prediction, the learning rate is 8e-4 with a batch size of 64.

\subsection{Evaluation Metrics}
For the protein function prediction task, Fmax and AUPR were used to evaluate the performance of models. Fmax measures the best harmonic mean of precision and recall over all decision thresholds. Specifically, for a given threshold $t$, the precision and recall are defined as
\begin{equation}
    \text{Pr(t)}=\frac{1}{f(t)}\sum_{i=1}^{f(t)}\frac{\sum_{j=1}^M\mathbb{I}(\hat{y_{i,j}}\geq t) y_{i,j}}{\sum_{j=1}^M\mathbb{I}(\hat{y_{i,j}}\geq t)}
\end{equation}
\begin{equation}
    \text{Rc(t)}=\frac{1}{N}\sum_{i=1}^N\frac{\sum_{j=1}^M\mathbb{I}(\hat{y_{i,j}}\geq t) y_{i,j}}{\sum_{j=1}^My_{i,j}}
\end{equation}
where $N$ denotes the number of proteins, $M$ denotes the number of functional labels, $y_{i,j}$ represents the ground-truth label, $\hat{y}_{i,j}$ is the predicted confidence score, $\mathbb{I}(\cdot)$ is the indicator function, and $f(t)$ denotes the number of proteins for which at least one function is predicted with confidence $\geq t$. The Fmax score is then computed as

\begin{equation}
    \text{Fmax}=\max_{\text{t}}\left\{\frac{2 \text{Pr(t)} \text{Rc(t)}}{\text{Pr(t)}+\text{Rc(t)}}\right\}
\end{equation}
where $f(t)$ is the number of proteins that predict at least one function with confidence $\geq t$. AUPR is the area under the precision-recall curve.

For the protein subcellular localization prediction task, the following metrics are used:
\begin{itemize}
    \item MicroF1: the F1 score computed by considering the total number of true positives, false negatives, and false positives.
    \item MacroF1: the averaged F1 score over all classes.
    \item Matthews correlation coefficient: a metric computed for each class, requiring the model to perform well with respect to all four entries of the confusion matrix.
\end{itemize}

For the mutation-induced protein solubility prediction task, the correct prediction ratio (CPR), generalized correlation (GC$^2$), and their normalized versions were used to evaluate model performance.
\begin{equation}
    \text{CPR}=\frac{1}{N}z_{i,i}
\end{equation}
\begin{equation}
    \text{GC}^2=\frac{1}{N(K-1)}\sum_{i,j}\frac{(z_{i,j}-e_{i,j})^2}{e_{i,j}}
\end{equation}
where $N$ is the sample number and $K$ is the class number. Here $z_{i,j}$ is the number of samples of class $i$ to class $j$. Let $x_i=\sum_jz_{i,j}$ be the number of inputs associated with class $i$, $y_j=\sum_iz_{i,j}$ be the number of samples predicted to be in class $i$, then 
\begin{equation}
    e_{i,j}=\frac{x_i\times y_j}{N}
\end{equation}

For the protein--protein binding affinity prediction task, Pearson correlation coefficient (PCC), Root Mean Square Error (RMSE), and Mean Absolute Error (MAE) were used to evaluate model performance. For the protein--protein binding binary classification task, accuracy was used as the evaluation metric.

For mutation-induced protein--protein binding affinity change prediction, Pearson correlation coefficient (PCC) was used as the evaluation metric. For cyclic peptide function prediction, the following metrics are used
\begin{equation}
    \text{Precision}=\frac{1}{n}\sum_{i=1}^n\frac{||L_i\cap L_i^*||}{L_i^*}
\end{equation}
\begin{equation}
    \text{Coverage}=\frac{1}{n}\sum_{i=1}^n\frac{||L_i\cap L_i^*||}{L_i}
\end{equation}
\begin{equation}
    \text{Accuracy}=\frac{1}{n}\sum_{i=1}^n\frac{||L_i\cap L_i^*||}{L_i\cup L_i^*}
\end{equation}
\begin{equation}
    \text{Absolute True}=\frac{1}{n}\sum_{i=1}^n\Delta(L_i,L_i^*)
\end{equation}
\begin{equation}
    \text{Absolute False}=\sum_{i=1}^n\frac{||L_i\cup L_i^*||-||L_i\cap L_i^*||}{m}
\end{equation}
where $n$ denotes the total number of samples, $m$ denotes the functional types of the cyclic peptides, $||\cdot||$ represents the cardinality of a set, $L_i$ is the subset of true labels for the $i$-th sample, $L_i^*$ is the subset of predicted labels for the $i$-th sample and
\begin{equation}
\Delta(L_i,L_i^*)=
\begin{cases}
1, & L_i=L_i^*,\\
0, & \text{otherwise}.
\end{cases}
\end{equation}

\subsection{Model Transferability to Diverse Protein Property Tasks}
For all downstream property prediction tasks, the AlphaFunctor architecture is adapted according to the input composition of each task. For example, a single AlphaFunctor model is used for single-protein tasks, whereas two AlphaFunctor models are used for double-protein tasks. The resulting embeddings are concatenated and passed through a two-layer multilayer perceptron (MLP) to generate the final prediction. During transfer learning, the entire model is fine-tuned in an end-to-end manner. The details are provided as follows.

For protein subcellular localization prediction, since subcellular localization labels largely correspond to a subset of the CC terms, the model trained on the CC task was adopted, and only the final linear layer was replaced to match the target localization classes. Five runs with different random seeds were performed and the average results were used as the final results of our model.

For mutation-induced protein solubility prediction, the wild-type and mutant protein sequences were processed separately by the pretrained model, and the resulting embeddings were concatenated and subsequently fed into a two-layer MLP for prediction. The models pretrained on the CC, BP, and MF tasks were each used independently. For each task-specific model, five runs with different random seeds were conducted, and the average of all $15 = 3 \times 5$ predictions was taken as the final output of our framework. 

For protein--protein binding affinity prediction, the amino acid sequences of the two binding partners were separately processed by the pretrained model, and the resulting embeddings were concatenated and fed into a two-layer MLP for prediction. Similar to the adaptation to solubility prediction, the models pretrained on the CC, BP, and MF tasks were each used independently. For each task-specific model, five runs with different random seeds were conducted, and the average of all $15 = 3 \times 5$ predictions was taken as the final output of our framework. On the S79 dataset, our model outperforms all existing methods in terms of PCC, RMSE, and MAE. Specifically, it achieves a PCC of 0.711, an RMSE of 1.96 kcal/mol, and an MAE of 1.46 kcal/mol, compared with the second-best model's score of 0.710, 2.00 kcal/mol, and 1.49 kcal/mol, respectively. Furthermore, when only the model trained on the BP task is used, AlphaFunctor achieves even better performance, with a PCC of 0.719, an RMSE of 1.93 kcal/mol, and an MAE of 1.45 kcal/mol.

For protein--protein interaction classification, the amino acid sequences of the two binding partners were processed separately using the pretrained model. The resulting embeddings were then concatenated and passed into a two-layer MLP for prediction. Similar to the adaptation strategy used for solubility prediction, the models pretrained on the CC, BP, and MF tasks were employed independently. For each task-specific model, we conducted five runs with different random seeds, and the average of all $15 = 3 \times 5$ predictions was used as the final output of our framework. On the Yeast-PPI dataset, our model achieved an ACC of 0.665. Notably, when using only the model pretrained on the CC task, AlphaFunctor achieved even better performance, with an ACC of 0.690.

For mutation-induced protein--protein binding affinity change prediction, the wild-type and mutant amino acid sequences of the two binding partners were processed separately by the pretrained model, and the resulting embeddings were concatenated and fed into a two-layer MLP for prediction. Based on the results of the protein--protein binding affinity prediction task, the model trained on the BP task was employed. Five runs with different random seeds were conducted, and the average prediction across these runs was used as the final output of our framework.

For cyclic peptide function prediction, the amino acid sequence of each cyclic peptide was processed by the pretrained model, and the resulting embeddings were fed into a two-layer MLP for prediction. The models pretrained on the CC, BP, and MF tasks were used independently. For each pretrained model, five runs with different random seeds were conducted, and the average of all $15=3\times 5$ predictions was taken as the final output of our framework.

\section{Robustness Analysis}
\subsection{Evaluation on the SwissProt Dataset for Protein Function Prediction}
The SwissProt dataset \cite{boutet2007uniprotkb} is currently the largest resource containing manually-annotated protein function records, and thus serves as a reliable benchmark for evaluating protein function prediction models. Since the long-tail distribution of the sequence lengths, we remove sequences longer than 5000 residues, resulting in a dataset of 552,711 protein sequences. To ensure a robust and unbiased analysis, the CD-HIT \cite{fu2012cd} is used to cluster all sequences at a 40\% sequence identity. This procedure results in 70,513 clusters for CC, 67,153 clusters for BP, and 66,728 clusters for MF. Then a greedy balanced matching strategy is used to partition these clusters into five groups, which are then used for five-fold cross-validation. We evaluate our AlphaFunctor model on these splits using five-fold cross-validation, the results are illustrated in Table \ref{tab:result-swiss-prot}.
\begin{table}[h]
    \centering
    \caption{AlphaFunctor performance in terms of Fmax and AUPR  for the Swiss-Prot dataset obtained using five-fold cross-validation. 
    }
    \begin{tabular}{l|cc|cc|cc}
    \hline
     & \multicolumn{2}{c|}{MF} & \multicolumn{2}{c|}{CC} & \multicolumn{2}{c}{BP}  \\
    & Fmax   & AUPR   & Fmax   & AUPR   & Fmax   & AUPR     \\
    \hline
   Fold 1 & 0.864	&0.905	&0.823	&0.898	&0.807	&0.859\\
   \hline
   Fold 2 & 0.877	&0.920	&0.827	&0.899	&0.791	&0.846\\
   \hline
   Fold 3 & 0.866	&0.901	&0.832	&0.900	&0.781	&0.838\\
   \hline
   Fold 4 & 0.877	&0.908	&0.814	&0.879	&0.804	&0.849\\
   \hline
   Fold 5 & 0.856	&0.901	&0.826	&0.904	&0.796	&0.841\\
   \hline
   Average & 0.868	&0.907&	0.824&	0.896&	0.796	&0.847 \\
    \hline
    \end{tabular}
    \label{tab:result-swiss-prot}
\end{table}
It can be seen that AlphaFunctor demonstrates strong and consistent performance across all three Gene Ontology categories. Specifically, for the averaged results, AlphaFunctor achieves an Fmax of 0.868 and an AUPR of 0.907 for MF, an Fmax of 0.824 and an AUPR of 0.896 for CC, and an Fmax of 0.796 and an AUPR of 0.847 for BP. The relatively lower performance on BP can be attributed to its substantially larger label space, comprising 17,511 functional terms, in contrast to 2,879 for CC and 8,110 for MF. Notably, AlphaFunctor performs best on the MF category despite CC having fewer labels, suggesting that factors beyond label cardinality, such as functional diversity and annotation complexity, also play an important role. Overall, these results indicate that AlphaFunctor achieves competitive performance on large-scale, high-quality datasets, highlighting its effectiveness for protein function prediction.

\subsection{Robustness with Respect to  Sequence Lengths and Species}
\begin{figure}[ht]
    \centering
    \includegraphics[width=1\linewidth]{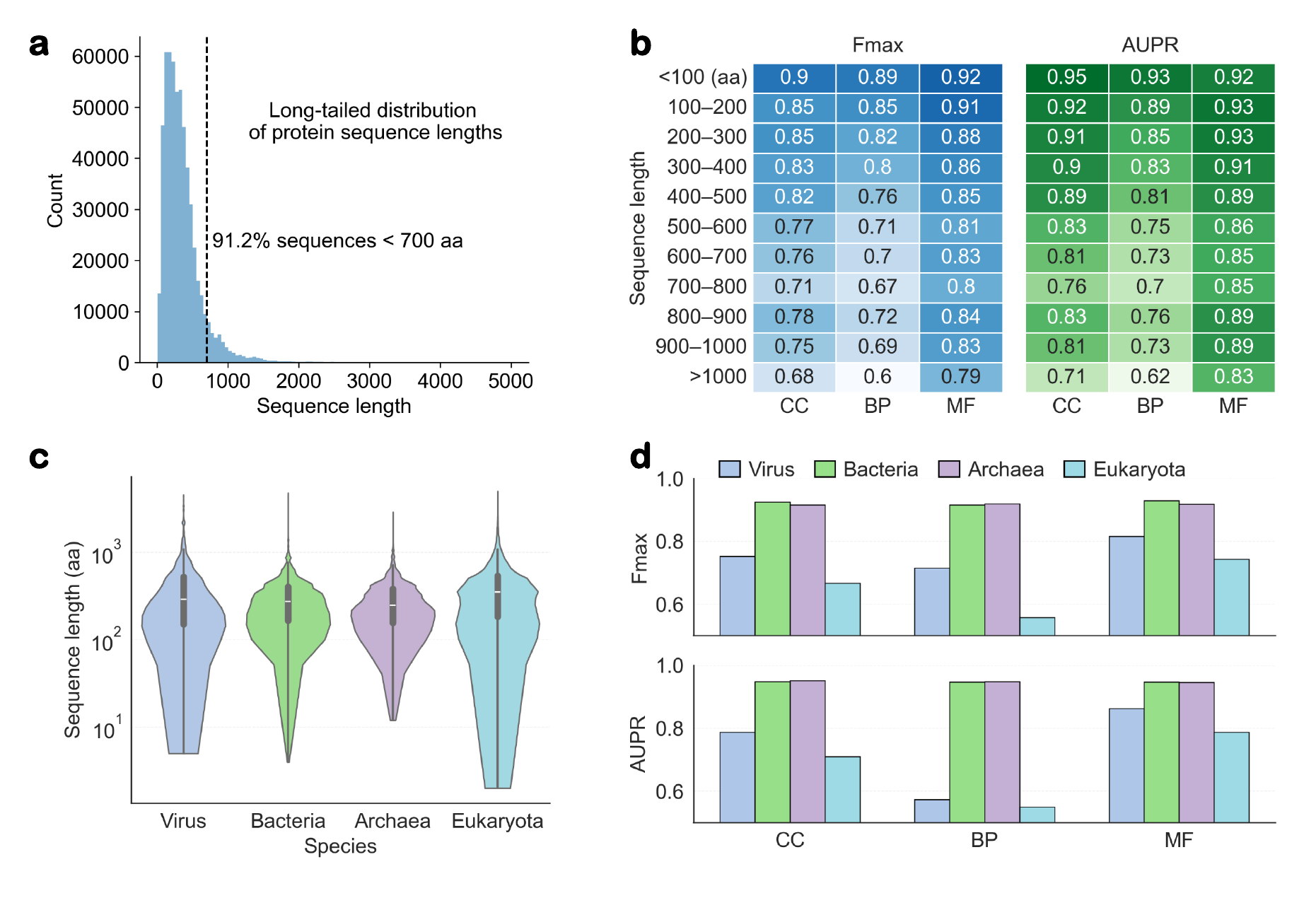}
    \caption{Model performance of AlphaFunctor across groups defined by sequence length and species. (a): Long-tailed distribution of sequence lengths in the Swiss-Prot dataset. (b): Model performance of AlphaFunctor across groups stratified by sequence length. (c): Sequence length distributions across different species, including virus, bacteria, archaea, and eukaryota. (d): Model performance of MVTDL across different species groups.}
    \label{fig:robust-length-species}
\end{figure}
To assess the robustness of AlphaFunctor, we stratify the SwissProt dataset according to sequence length and species, and evaluate model performance within each group using predictions obtained from five-fold cross-validation. The results are presented in Figure \ref{fig:robust-length-species}.

As shown in Figure \ref{fig:robust-length-species}a, protein sequence lengths exhibit a clear long-tailed distribution. Specifically, 96.7\% of sequences are shorter than 1,000 amino acids, and 80.4\% are shorter than 500 amino acids, whereas only 0.04\% (230 sequences) exceed 4,000 amino acids. This long-tailed distribution introduces challenges for protein function prediction due to the imbalance across different length ranges. 
To further investigate this effect, we partition the dataset into 11 length intervals: 0--100, 100--200, 200--300, 300--400, 400--500, 500--600, 600--700, 700--800, 800--900, 900--1000, and 1000--5000. The corresponding performance of AlphaFunctor is shown in Figure \ref{fig:robust-length-species}b. Across all three tasks, a consistent trend is observed where performance gradually decreases as sequence length increases.
Notably, AlphaFunctor achieves consistently strong performance on the MF task, with an Fmax of at least 0.79 and an AUPR of at least 0.83 across all groups, and averaged Fmax of 0.85 and AUPR of 0.89. In comparison, the CC task shows slightly lower performance, with average Fmax and AUPR values of 0.79 and 0.85, respectively, while the BP task yields the lowest performance, with an average Fmax of 0.75 and AUPR of 0.78. The relatively lower performance on the BP task can be attributed to the larger number and higher complexity of functional terms within this category.
Overall, AlphaFunctor achieves at least an average Fmax of 0.75 and an average AUPR of 0.78 across all tasks and length groups, demonstrating its effectiveness in protein function prediction.
These results also suggest that AlphaFunctor is more capable on the MF task compared to CC and BP. 
Focusing on sequences shorter than 700 amino acids, which account for 91.2\% of the dataset, AlphaFunctor obtains average Fmax/AUPR values of 0.83/0.89 for CC, 0.79/0.83 for BP, and 0.87/0.90 for MF. Furthermore, for the shorter sequence groups, including 0--100, 100--200, 200--300, and 300--400, which together constitute a substantial proportion of the dataset, that is 9.9\%, 21.2\%, 19.7\%, and 17.5\%, respectively, AlphaFunctor consistently achieves at least 0.82 Fmax and 0.85 AUPR across all three tasks. These results highlight the robustness and effectiveness of AlphaFunctor across the majority of protein sequences, particularly within the dominant short-to-medium length ranges.

We further divide the protein sequences into four groups according to species, including virus, bacteria, archaea, and eukaryota. The sequence length distributions of these groups are presented in Figure \ref{fig:robust-length-species}c, and the corresponding model performance is shown in Figure \ref{fig:robust-length-species}d.
Notably, AlphaFunctor achieves consistently strong performance on the bacteria and archaea groups across all three tasks. Specifically, for the bacteria group, the Fmax scores are 0.925, 0.915, and 0.929, with corresponding AUPR values of 0.948, 0.947, and 0.947 for CC, BP, and MF, respectively. For the archaea group, the Fmax scores are 0.915, 0.919, and 0.918, while the AUPR values are 0.952, 0.948, and 0.946 for CC, BP, and MF, respectively. This strong performance can be partially attributed to the relatively concentrated sequence length distributions of bacteria and archaea, which are centered around shorter sequence lengths of approximately 300 amino acids. In contrast, the virus group exhibits a broader range of sequence lengths, while eukaryotic proteins span the widest range, which may increase the complexity of sequence--function relationships and lead to comparatively lower performance. The strong performance on bacteria and archaea suggests that AlphaFunctor is well suited for large-scale functional annotation of prokaryotic proteomes, with potential applications in microbiome analysis and antibiotic target discovery. 
Although the performance on the virus group is relatively lower, AlphaFunctor still achieves an average Fmax of 0.761 and an average AUPR of 0.741 across the three functional categories. This demonstrates its capability in viral protein annotation, which is particularly important for emerging viruses, where rapid and accurate functional characterization is critical for understanding pathogenic mechanisms and facilitating vaccine development.

\subsection{Robustness with respect to Sequence Identity}
\begin{figure}
    \centering
    \includegraphics[width=1\linewidth]{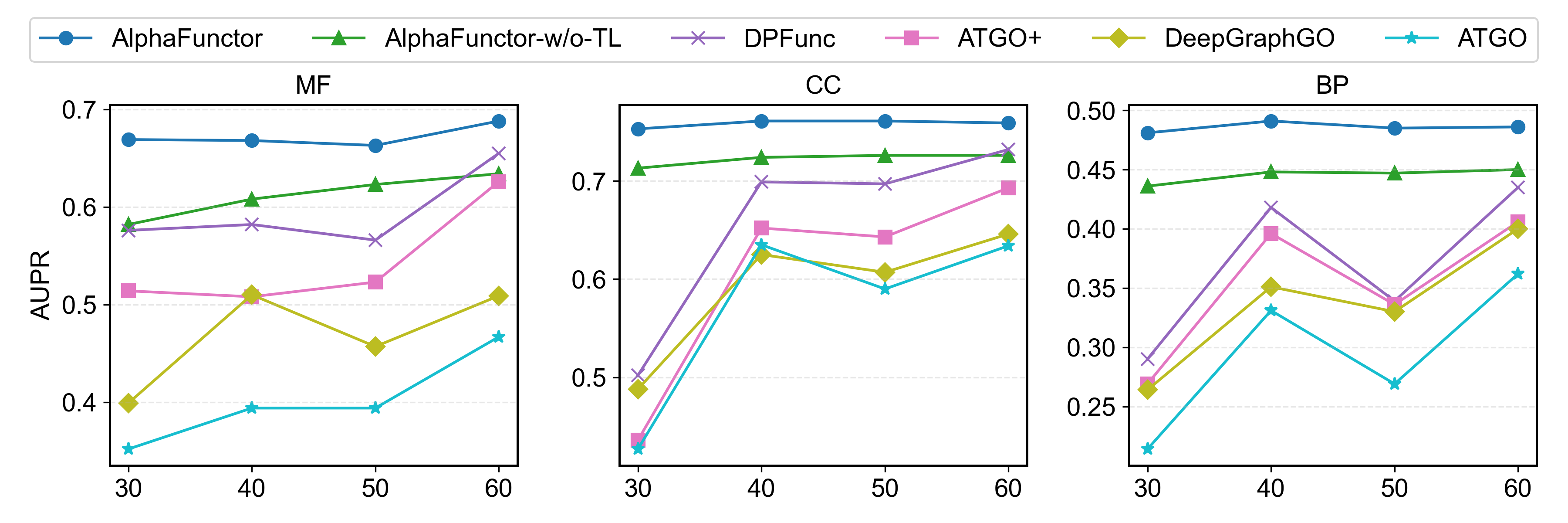}
    \caption{Ablation study of AlphaFunctor over sequence identity. AlphaFunctor and AlphaFunctor-w/o-TL denote our models with and without transfer learning, respectively. The results show that AlphaFunctor consistently outperforms existing methods across all sequence identity levels and tasks. Moreover, both AlphaFunctor and AlphaFunctor-w/o-TL exhibit stable performance across varying sequence identity ranges.}
    \label{fig:sequence-similarity}
\end{figure}

The homology between protein sequences may result in highly similar samples appearing in both the training and testing sets, potentially leading to data leakage and an overestimation of model performance. To mitigate this issue and provide a more reliable evaluation of model robustness and generalization, we partition the data based on different sequence identity thresholds. The existing best method DPFunc \cite{wang2025dpfunc} has conducted a similar analysis on the CAFA4 dataset. Following the same protocol, we construct multiple subsets of the test data, where each subset contains proteins sharing no more than a specified sequence identity threshold with any protein in the training set. These subsets are generated using MMseqs2 \cite{steinegger2017mmseqs2}. The considered identity thresholds are 30\%, 40\%, 50\%, and 60\%, and the corresponding results are presented in Figure \ref{fig:sequence-similarity}. 
We evaluate both AlphaFunctor with transfer learning and AlphaFunctor without transfer learning, and compare them with four leading baseline methods. As shown in Figure \ref{fig:sequence-similarity}, AlphaFunctor consistently outperforms existing approaches across all three tasks and all sequence identity thresholds. Notably, both AlphaFunctor and AlphaFunctor-no-TL maintain stable performance as the sequence identity decreases. Even at the most challenging threshold of 30\%, our model achieves strong performance across all tasks. These results demonstrate the robustness of our model and its strong lower bound performance under varying levels of sequence similarity.

\subsection{Robustness with Respect to Ground Truth Dataset}
\begin{figure}
    \includegraphics[width=0.5\linewidth]{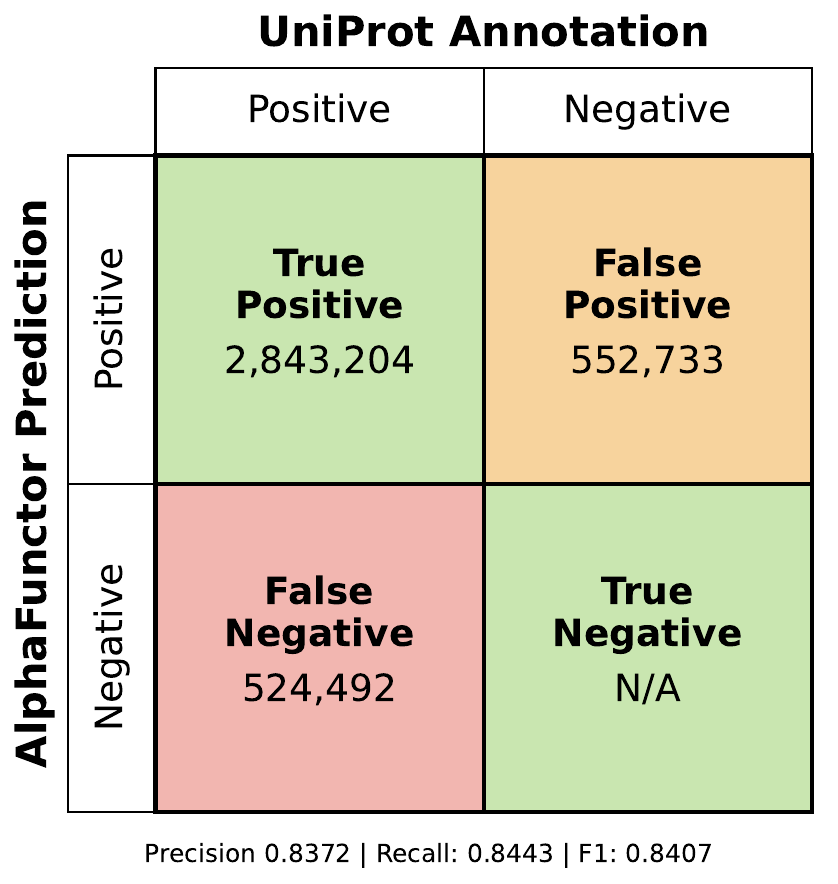}\includegraphics[width=0.5\linewidth]{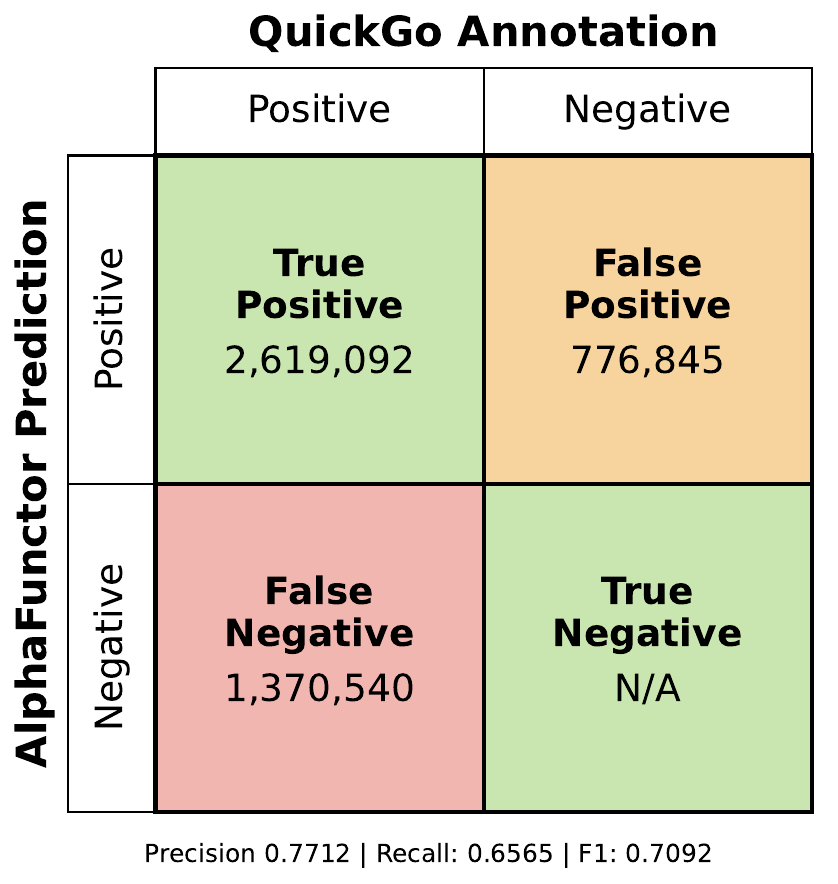}
\caption{\label{fig:quickgocompare}Comparison for the confusion matrices depending on the ground truth reference to either UniProt or QuickGO for all proteins in SwissProt.
}
\end{figure}

AlphaFunctor was originally trained using the SwissProt annotations contained in UniProt, which excludes most automatic or redundant annotations.
From Figure~\ref{fig:sequence-similarity}, we find that evaluating the AlphaFunctor predictions using ground truth based on UniProt noticably improves prediction performance.
Rather than strictly being a superset, QuickGO contains both additional annotations determined automatically, and updates legacy GO terms with more regularity than UniProt does.
Thus, many of our legacy GO term predictions end up in the false positive bucket in Figure~\ref{fig:quickgocompare}.

Likewise, without explicitly training on automatic annotations, AlphaFunctor is missing many automatically annotated GO terms.
This results in a higher rate of false negatives when comparing to QuickGO.
If matching QuickGO outputs was the ultimate goal, retraining AlphaFunctor based on this larger annotation base would be required.
However, this may ultimately lower the quality of the GO term assignments in the training data, muddying its predictive power when applied to novel proteins well outside of the training set.
At this juncture, we feel it is unclear which approach is ultimately superior, and have used the more conservative dataset as the ground truth thoughout the rest of the article.

\section{Supplementary Tables}
Tables \ref{tab:result-cafa3}--\ref{tab:result-cyclic-peptide} summarize the performance comparisons between AlphaFunctor and existing methods across protein function annotation and downstream protein property prediction tasks. Specifically, Tables \ref{tab:result-cafa3}, \ref{tab:result-cafa4}, and \ref{tab:result-pdb} report the Fmax and AUPR results on the CAFA3, CAFA4, and PDB datasets for protein function annotation, respectively. Table \ref{tab:swissprot_cv} presents the five-fold cross-validation results on the SwissProt dataset for subcellular localization prediction, while Table \ref{tab:slp-result} reports the results on the independent HPA test set. Tables \ref{tab:result-solubility}, \ref{tab:result-s79}, \ref{tab:result-yeast}, \ref{tab:result-skempi}, and \ref{tab:result-cyclic-peptide} further provide comparisons on the PON-Sol2 dataset for mutation-induced protein solubility prediction, the S79 dataset for protein--protein binding affinity prediction, the Yeast-PPI dataset for protein--protein interaction classification, the SKEMPI dataset for mutation-induced protein--protein binding affinity change prediction, and the CyclicPepedia dataset for cyclic peptide function prediction, respectively.

\begin{table*}[htp]
\centering
\caption{Performance comparison on the CAFA3 dataset. The benchmark results are taken from the corresponding publications 
\cite{kulmanov2018deepgo,kulmanov2020deepgoplus,cao2021tale,wang2023mmsmaplus,song2024deepss2go,yang2026multi}}
\label{tab:result-cafa3}
\begin{tabular}{l|cc|cc|cc}
\hline
\multirow{2}{*}{Method} & \multicolumn{2}{c|}{MF (677 labels)} & \multicolumn{2}{c|}{CC (551 labels)} & \multicolumn{2}{c}{BP (3992 labels)} \\
 & Fmax & AUPR & Fmax & AUPR & Fmax & AUPR \\
\hline
Naive        & 0.290 & 0.130 & 0.562 & 0.456 & 0.357 & 0.254 \\
DiamondBLAST & 0.431 & 0.178 & 0.506 & 0.142 & 0.399 & 0.116 \\
DiamondScore & 0.509 & 0.340 & 0.557 & 0.335 & 0.427 & 0.267 \\
DeepGO       & 0.393 & 0.303 & 0.565 & 0.579 & 0.435 & 0.385 \\
DeepGOCNN    & 0.420 & 0.355 & 0.607 & 0.616 & 0.378 & 0.323 \\
MMSMA        & 0.583 & 0.541 & 0.620 & 0.590 & 0.518 & 0.457 \\
TALE+        & 0.558 & 0.539 & 0.622 & 0.595 & 0.480 & 0.427 \\
DeepGOPlus   & 0.544 & 0.487 & 0.623 & 0.627 & 0.469 & 0.404 \\
MMSMAPlus    & 0.595 & 0.559 & 0.622 & 0.601 & 0.535 & 0.470 \\
DeepSS2GO    & 0.601 & 0.559 & 0.643 & 0.634 & 0.518 & 0.441 \\
MVCFF	     & 0.597 & 0.568 & 0.631 & 0.634 & 0.537 & 0.469 \\
AlphaFunctor-w/o-TL& 0.673 & 0.730 & 0.703 & 0.766 & 0.589 & 0.596 \\
AlphaFunctor        & 0.698 & 0.756 & 0.716 & 0.780 & 0.599 & 0.581 \\
\hline
\end{tabular}
\end{table*}

\begin{table*}[htp]
\centering
\caption{Performance comparison on the CAFA4 dataset. The benchmark results are taken from the corresponding publications \cite{radivojac2013large,buchfink2015fast,kulmanov2018deepgo,kulmanov2020deepgoplus,cao2021tale,you2021deepgraphgo,zhu2022integrating,wang2025dpfunc}}
\label{tab:result-cafa4}
\begin{tabular}{l|cc|cc|cc}
\hline
\multirow{2}{*}{Method} & \multicolumn{2}{c|}{MF (6166 labels)} & \multicolumn{2}{c|}{CC (2548 labels)} & \multicolumn{2}{c}{BP (19825 labels)} \\
 & Fmax & AUPR & Fmax & AUPR & Fmax & AUPR \\
\hline
Diamond      & 0.592 & 0.387 & 0.573 & 0.283 & 0.429 & 0.197 \\
BlastKNN     & 0.616 & 0.484 & 0.596 & 0.384 & 0.445 & 0.258 \\
DeepGO       & 0.301 & 0.204 & 0.574 & 0.580 & 0.328 & 0.260 \\
DeepGOCNN    & 0.396 & 0.326 & 0.573 & 0.567 & 0.323 & 0.254 \\
TALE         & 0.260 & 0.158 & 0.548 & 0.510 & 0.253 & 0.152 \\
ATGO         & 0.454 & 0.442 & 0.602 & 0.592 & 0.396 & 0.341 \\
DeepGraphGO  & 0.562 & 0.533 & 0.634 & 0.590 & 0.432 & 0.389 \\
DeepGOPlus   & 0.589 & 0.548 & 0.626 & 0.618 & 0.438 & 0.365 \\
TALE+        & 0.602 & 0.543 & 0.608 & 0.591 & 0.427 & 0.327 \\
ATGO+        & 0.622 & 0.599 & 0.633 & 0.636 & 0.456 & 0.399 \\
DPFunc       & 0.635 & 0.658 & 0.657 & 0.695 & 0.466 & 0.434 \\
AlphaFunctor-w/o-TL& 0.692 & 0.722 & 0.690 & 0.744 & 0.494 & 0.498 \\
AlphaFunctor        & 0.709 & 0.762 & 0.702 & 0.763 & 0.517 & 0.516 \\
\hline
\end{tabular}
\end{table*}

\begin{table*}[htp]
\centering
\caption{Performance comparison on the PDB dataset. The benchmark results are taken from the corresponding publications \cite{radivojac2013large,kulmanov2018deepgo,kulmanov2020deepgoplus,gligorijevic2021structure,lai2022accurate,wang2025dpfunc,lin2025gobeacon}}
\label{tab:result-pdb}
\begin{tabular}{l|cc|cc|cc}
\hline
\multirow{2}{*}{Method} & \multicolumn{2}{c|}{MF (489 labels)} & \multicolumn{2}{c|}{CC (320 labels)} & \multicolumn{2}{c}{BP (1943 labels)} \\
 & Fmax & AUPR & Fmax & AUPR & Fmax & AUPR \\
\hline
Na\"{\i}ve   & 0.156 & 0.075 & 0.318 & 0.158 & 0.244 & 0.131 \\
BLAST        & 0.498 & 0.120 & 0.398 & 0.163 & 0.400 & 0.120 \\
DeepGO       & 0.359 & 0.368 & 0.420 & 0.302 & 0.295 & 0.210 \\
DeepFRI      & 0.542 & 0.313 & 0.424 & 0.193 & 0.425 & 0.159 \\
GAT-GO       & 0.633 & 0.660 & 0.547 & 0.479 & 0.492 & 0.381 \\
GOBeacon     & 0.617 & ---   & 0.579 & ---   & 0.484 & ---   \\
DPFunc       & 0.681 & 0.701 & 0.571 & 0.593 & 0.531 & 0.540 \\
AlphaFunctor-w/o-TL& 0.705 & 0.755 & 0.603 & 0.619 & 0.557 & 0.590 \\
AlphaFunctor        & 0.733 & 0.786 & 0.654 & 0.696 & 0.581 & 0.623 \\
\hline
\end{tabular}
\end{table*}

\begin{table}[htbp]
\centering
\caption{Performance comparison between our model, AlphaFunctor, and existing methods on the SwissProt dataset using five-fold cross-validation. The existing results are taken from the publication \cite{thumuluri2022deeploc}. The best results are in bold.}
\label{tab:swissprot_cv}
\resizebox{\textwidth}{!}{
\begin{tabular}{lcccccc}
\toprule
 & Counts & DeepLoc 1.0$^{\beta}$ & YLoc+$^{\alpha}$ 
 & \multicolumn{2}{c}{DeepLoc 2.0} & AlphaFunctor \\
\cmidrule(lr){5-6}
 &  &  &  & ESM1b & ProtT5 &  \\
\midrule
Type &  & Single & Multi & Multi & Multi & Multi \\


MicroF1 
& 28303 & $0.58 \pm 0.02$ & $0.56 \pm 0.01$ 
& $0.72 \pm 0.01$ & $0.73 \pm 0.01$ & $\mathbf{0.81\pm0.01}$ \\

MacroF1 
& 28303 & $0.47 \pm 0.01$ & $0.42 \pm 0.01$ 
& $0.64 \pm 0.01$ & $0.66 \pm 0.01$ & $\mathbf{0.77\pm0.01}$ \\

\midrule
\multicolumn{7}{l}{MCC per location ($\uparrow$ is better)} \\

Cytoplasm 
& 9870 & $0.45 \pm 0.02$ & $0.38 \pm 0.02$ 
& $0.61 \pm 0.01$ & $0.62 \pm 0.01$ & $\mathbf{0.69\pm0.01}$ \\

Nucleus 
& 9720 & $0.46 \pm 0.02$ & $0.42 \pm 0.02$ 
& $0.66 \pm 0.02$ & $0.69 \pm 0.01$ & $\mathbf{0.78\pm0.01}$ \\

Extracellular 
& 3301 & $0.78 \pm 0.05$ & $0.61 \pm 0.05$ 
& $0.85 \pm 0.03$ & $0.85 \pm 0.04$ & $\mathbf{0.88\pm0.02}$ \\

Cell membrane 
& 4187 & $0.53 \pm 0.02$ & $0.44 \pm 0.02$ 
& $0.64 \pm 0.01$ & $0.66 \pm 0.01$ & $\mathbf{0.73\pm0.02}$ \\

Mitochondrion 
& 2590 & $0.58 \pm 0.04$ & $0.47 \pm 0.02$ 
& $0.73 \pm 0.03$ & $0.76 \pm 0.02$ & $\mathbf{0.81\pm0.03}$ \\

Plastid 
& 1047 & $0.69 \pm 0.04$ & $0.72 \pm 0.02$ 
& $0.88 \pm 0.01$ & $0.90 \pm 0.01$ & $\mathbf{0.91\pm0.02}$ \\

Endoplasmic reticulum 
& 2180 & $0.32 \pm 0.04$ & $0.17 \pm 0.04$ 
& $0.52 \pm 0.01$ & $0.56 \pm 0.03$ & $\mathbf{0.69\pm0.02}$ \\

Lysosome/Vacuole 
& 1496 & $0.06 \pm 0.05$ & $0.07 \pm 0.03$ 
& $0.24 \pm 0.03$ & $0.28 \pm 0.04$ & $\mathbf{0.56\pm0.03}$ \\

Golgi apparatus 
& 1279 & $0.20 \pm 0.04$ & $0.11 \pm 0.04$ 
& $0.36 \pm 0.06$ & $0.34 \pm 0.05$ & $\mathbf{0.62\pm0.03}$ \\

Peroxisome 
& 304 & $0.15 \pm 0.04$ & $0.05 \pm 0.02$ 
& $0.48 \pm 0.05$ & $0.56 \pm 0.08$ & $\mathbf{0.72\pm0.05}$ \\

\bottomrule
\end{tabular}
}
\end{table}

\begin{table}[hbp]
\centering
\caption{Performance comparison between AlphaFunctor and existing methods on the HPA independent test set for subcellular localization prediction. The existing results are extracted from the corresponding publication~\cite{thumuluri2022deeploc}. The best results are in bold.}
\label{tab:slp-result}
\resizebox{\textwidth}{!}{
\begin{tabular}{l |c |c |c |c |c |c c |c}
\toprule
 & Count & YLoc+ & DeepLoc 1.0 & Fuel-mLoc & LAProtT5 & \multicolumn{2}{c|}{DeepLoc 2.0} & AlphaFunctor\\
 &  & Animal &  & Euk &  & ESM1b & ProtT5 &  \\
\midrule
MicroF1 & 1717 & 0.51 & 0.46 & 0.52 & 0.56 & 0.57 & 0.60 &\textbf{0.64}\\
MacroF1 & 1717 & 0.34 & 0.35 & 0.39 & 0.43 & 0.44 & 0.46&\textbf{0.51} \\

\midrule
\multicolumn{8}{l}{MCC per location ($\uparrow$ is better)} \\

Cytoplasm & 562 & 0.14 & 0.23 & 0.23 & 0.33 & 0.29 & 0.36 &\textbf{0.39}\\
Nucleus & 893 & 0.20 & 0.28 & 0.41 & 0.45 & 0.41 & 0.44& \textbf{0.52} \\
Cell membrane & 287 & 0.20 & 0.23 & 0.32 & 0.30 & 0.34 & 0.36 & \textbf{0.43}\\
Mitochondrion & 196 & 0.37 & 0.39 & 0.33 & 0.59 & 0.60 & 0.56 &\textbf{0.65} \\
Endoplasmic reticulum & 77 & 0.12 & \textbf{0.23} & 0.14 & 0.22 & 0.20 & 0.17 & 0.15\\
Golgi apparatus & 86 & 0.08 & 0.10 & 0.24 & 0.26 & 0.17 & 0.31 & \textbf{0.47}\\
\bottomrule
\end{tabular}
}
\end{table}

\begin{table}[htbp]
\centering
\caption{Performance comparison between AlphaFunctor and existing methods on the PON-Sol2 dataset for mutation-induced protein solubility prediction. The existing results are extracted from the corresponding publication~\cite{wee2024integration}. The best results are in bold.}
\label{tab:result-solubility}
\begin{tabular}{lcccc}
\toprule
\multirow{2}{*}{Model} & \multicolumn{2}{c}{CPR} & \multicolumn{2}{c}{GC$^2$} \\
 & Unnorm. & Norm. & Unnorm. & Norm. \\
\midrule
SODA & 0.282 & 0.247 & NaN & NaN \\
PON-Sol & 0.356 & 0.389 & 0.010 & 0.011 \\
SODA (10 as Threshold) & 0.375 & 0.347 & 0.022 & 0.022 \\
SODA (5 as Threshold) & 0.381 & 0.341 & 0.041 & 0.045 \\
SODA (17 as Threshold) & 0.382 & 0.356 & 0.016 & 0.016 \\
PON-Sol2 & 0.671 & 0.545 & 0.181 & 0.157 \\
TopGBT & 0.707 & 0.562 & 0.205 & 0.184 \\
AlphaFunctor & \textbf{0.725} & \textbf{0.578} & \textbf{0.224} & \textbf{0.202} \\
\bottomrule
\end{tabular}
\end{table}

\begin{table}[htbp]
\centering
\caption{Performance comparison on S79 for protein-protein binding affinity prediction. The existing results are extracted from the corresponding publication~\cite{xu2025ssif}. The best results are in bold.}
\label{tab:result-s79}
\begin{tabular}{lccc}
\hline
\multirow{2}{*}{Method} & \multicolumn{3}{c}{S79} \\
 & PCC & MAE & RMSE   \\
\hline
DFIRE          & 0.602 & 4.64 & 5.37   \\
CP\_PIE        & 0.517 & 8.80 & 9.18    \\
ISLAND         & 0.378 & 2.10 & 2.65   \\
PPI-Affinity  & 0.616 & 1.82 & 2.23    \\
ProAffinity-GNN & 0.697 & 1.52 & 2.04  \\
PPI-Graphomer & 0.622 & 1.69 & 2.24   \\
SSIF-Affinity        & 0.710 & 1.49 & 2.00  \\
AlphaFunctor & \textbf{0.719} & \textbf{1.45} & \textbf{1.93}  \\
\hline
\end{tabular}
\end{table}

\begin{table}[htbp]
\centering
\caption{Performance (accuracy) on Yeast-PPI dataset for protein-protein interaction prediction. The benchmark results are taken from the publication \cite{ullanat2026learning}. The best results are in bold. }
\label{tab:result-yeast}
\begin{tabular}{lll}
\hline
Model & Concatenation Type & Yeast PPI \\
\hline
ProtT5-BFD & Concatenate tokens & 0.566 \\
ProtT5-UniRef & Concatenate tokens & 0.580 \\
Progen2-Large & Concatenate tokens & 0.578  \\
ESM-2-3B & Concatenate tokens & 0.598 \\
ESM-1b-650M & Concatenate tokens & 0.596  \\
ESM-2-650M & Concatenate tokens & 0.624 \\
ESM-2-150M & Concatenate tokens & 0.586 \\
\hline
ProtT5-BFD & Concatenate embeddings & 0.613  \\
ProtT5-UniRef & Concatenate embeddings & 0.588 \\
Progen2-Large & Concatenate embeddings & 0.584 \\
ESM-2-3B & Concatenate embeddings & 0.622  \\
ESM-1b-650M & Concatenate embeddings & 0.607 \\
ESM-2-650M & Concatenate embeddings & 0.607  \\
ESM-2-150M & Concatenate embeddings & 0.597 \\
\hline
MINT & Concatenate embeddings & 0.687 \\
\hline
AlphaFunctor & Concatenate embeddings & \textbf{0.690}\\
\hline
\end{tabular}
\end{table}

\begin{table}[htbp]
\centering
\caption{Performance (PCC) on the SKEMPI dataset for mutation-induced protein--protein binding affinity change prediction. The benchmark results are taken from the publication \cite{ullanat2026learning}. The best results are in bold.}
\label{tab:result-skempi}
\begin{tabular}{l l c}
\hline
Model & Concatenation Type & SKEMPI \\
\hline
ProtT5-BFD     & Concatenate tokens      & 0.295  \\
ProtT5-Uniref  & Concatenate tokens      & 0.316  \\
Progen2-Large  & Concatenate tokens      & 0.381  \\
ESM-2-3B       & Concatenate tokens      & 0.340  \\
ESM-1b-650M    & Concatenate tokens      & 0.278  \\
ESM-2-650M     & Concatenate tokens      & 0.227  \\
ESM-2-150M     & Concatenate tokens      & 0.319  \\
\hline
ProtT5-BFD     & Concatenate embeddings  & 0.270  \\
ProtT5-Uniref  & Concatenate embeddings  & 0.224  \\
Progen2-Large  & Concatenate embeddings  & 0.304  \\
ESM-2-3B       & Concatenate embeddings  & 0.231  \\
ESM-1b-650M    & Concatenate embeddings  & 0.251 \\
ESM-2-650M     & Concatenate embeddings  & 0.256  \\
ESM-2-150M     & Concatenate embeddings  & 0.292 \\
\hline
MINT           & Concatenate embeddings  & 0.403 \\
AlphaFunctor   & Concatenate embeddings  & \textbf{0.455} \\
\hline
\end{tabular}
\label{tab:skempi_results}
\end{table}

\begin{table}[htbp]
\centering
\caption{Performance on the CyclicPepedia dataset \cite{lu2026mfcpedpred} for cyclic peptide function prediction. The benchmark results are taken from the publication. The best results are in bold.}
\label{tab:result-cyclic-peptide}
\begin{tabular}{lccccc}
\hline
Method & Precision$\uparrow$ & Coverage$\uparrow$ & Accuracy$\uparrow$ & Absolute true$\uparrow$ & Absolute false$\downarrow$ \\
\hline
ChemBERTa-77M-MLM & 0.769 & 0.755 &0.697 &0.488  &0.079 \\
SMILES\_tokenized\_PubC &0.749 &0.735& 0.670& 0.449& 0.090\\
ChemBERTa-zinc-base-v1 &0.727 &0.703& 0.643 &0.421& 0.093\\
Multi\_CycGT     & 0.750 & 0.708 & 0.668 & 0.477 & 0.075 \\
MultiCycPermea   & 0.762 & 0.762 & 0.691 & 0.468 & 0.082 \\
MFCPepPred       & 0.805 & 0.811 & 0.748 & 0.547 & 0.067 \\
AlphaFunctor & \textbf{0.807} & \textbf{0.813} & \textbf{0.751} & \textbf{0.554} & \textbf{0.064}\\
\hline
\end{tabular}
\end{table}

\section{Supplementary Figures}

\begin{figure}[ht!]
    \centering
    \includegraphics[width=0.7\linewidth]{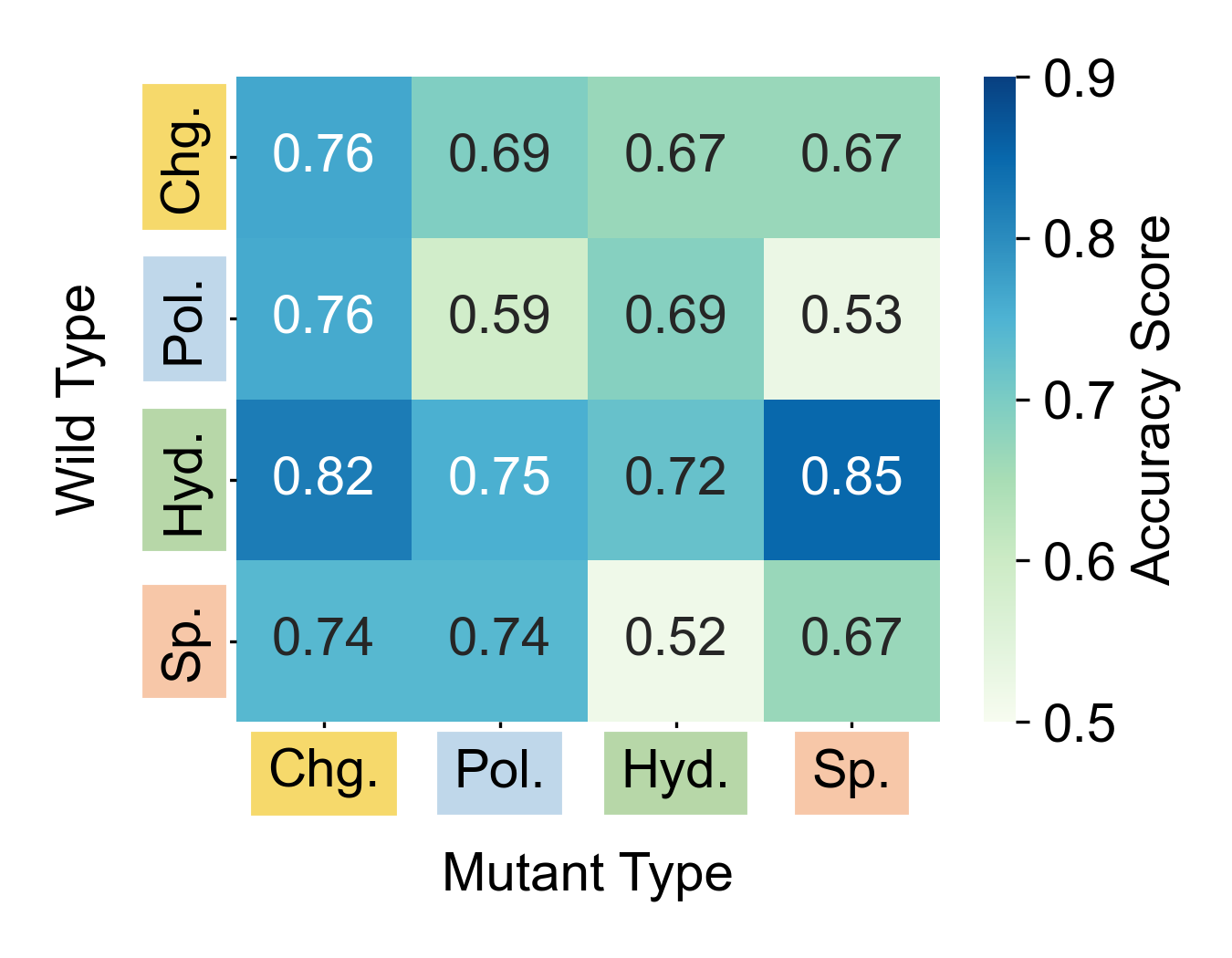}
    \caption{Model performance on different amino acid side-chain group type combinations on the PON-Sol dataset for mutation-induced protein solubility prediction.}
    \label{fig:placeholder}
\end{figure}

Figure \ref{fig:placeholder} shows the detailed accuracy of our model on different combinations of mutant-wide 
amino acid side-chain group types for the PON-Sol dataset. Notably, the combination of hydrophobic and charged group types has the highest predictive accuracy, while the combination of special and special group types has the lowest predictive accuracy.       

\pagebreak


\end{document}